\documentclass[journal,10pt]{IEEEtran}
\usepackage{amsfonts}
\IEEEoverridecommandlockouts

\ifCLASSINFOpdf
\else
\fi
\usepackage{epsfig}
\usepackage{graphicx}
\usepackage{psfig}
\usepackage{epsf}
\usepackage[cmex10]{amsmath}
\usepackage{booktabs}
\usepackage{color}
\usepackage{fancyhdr}
\usepackage{slashbox}
\usepackage{caption2}   

\begin{document}
\title{Artificial Noise Aided Secure Cognitive Beamforming for Cooperative MISO-NOMA Using SWIPT}
\author{\IEEEauthorblockN{Fuhui Zhou, \emph{Member, IEEE}, Zheng Chu, \emph{Member, IEEE}, Haijian Sun, \emph{Student Member, IEEE},\\
Rose Qingyang Hu, \emph{Senior Member, IEEE}, and Lajos Hanzo, \emph{Fellow, IEEE} }

\thanks{
Manuscript received September 15, 2017; revised Feb. 1, 2018 and
accepted February 15, 2018. Date of publication ****; date of
current version ****. The research of F. Zhou was supported in part by the Natural Science Foundation of China under Grant 61701214, in part by the Young Natural Science Foundation of Jiangxi Province under Grant 20171BAB212002, in part by The Open Foundation of The State Key Laboratory of Integrated Services Networks under Grant ISN19-08, and in part by The Postdoctoral Science Foundation of Jiangxi Province under Grant 2017M610400, Grant 2017KY04 and Grant 2017RC17. The research of Prof. R. Q. Hu was supported in  part  by  the US National Science Foundation grants ECCS-1308006 and EARS-1547312. The research of Prof. L. Hanzo was supported by the ERC for his Advanced Fellow Grant and the Royal Society for his Research Merit Award. }

\thanks{F. Zhou is with the Department of Electrical and Computer Engineering as a Research Fellow at Utah State University, U.S.A. F. Zhou is also with the School of Information Engineering, Nanchang University, P. R. China, 330031. He is also with State Key Laboratory of Integrated Services Networks, Xidian University, Xi¡¯an,
710071, P. R. China (e-mail: zhoufuhui@ieee.org).

Z. Chu is with 5G Innovation Center (5GIC), Institute for Communication Systems (ICS), University of Surrey, Guildford GU2 7XH, U.K (e-mail: z.chu@mdx.ac.uk).

H. Sun and R. Q. Hu are with Electrical and Computer Engineering Department, Utah State University, USA (e-mail: h.j.sun@ieee.org, rose.hu@usu.edu).

L. Hanzo is with the University of Southampton, Southampton SO17 1BJ, U.K. (e-mail: lh@ecs.soton.ac.uk).
}}
\maketitle
\begin{abstract}
Cognitive radio (CR) and non-orthogonal multiple access (NOMA) have been deemed two promising technologies due to their potential to achieve high spectral efficiency and massive connectivity. This paper studies a  multiple-input single-output NOMA CR network relying on simultaneous wireless information and power transfer (SWIPT) conceived for supporting a massive population of power limited battery-driven devices. In contrast to most of the existing works, which use an ideally linear energy harvesting model, this study applies a more practical non-linear energy harvesting model. In order to improve the security of the primary network, an artificial-noise-aided cooperative jamming scheme is proposed. The artificial-noise-aided beamforming design problems are investigated subject to the practical secrecy rate and energy harvesting constraints. Specifically, the transmission power minimization problems are formulated under both perfect channel state information (CSI) and  the bounded CSI error model. The problems formulated are non-convex, hence they are challenging to solve. A pair of algorithms either using semidefinite relaxation (SDR) or  a cost function are proposed for solving these problems. Our simulation results show that the proposed cooperative jamming scheme succeeds in establishing secure communications and NOMA is capable of outperforming the conventional orthogonal multiple access in terms of its power efficiency. Finally, we demonstrate that  the cost function  algorithm outperforms the SDR-based algorithm.
\end{abstract}
\IEEEpeerreviewmaketitle
\section{Introduction}
\IEEEPARstart{T}{HE} next generation wireless communication systems call for advanced communication techniques that can achieve high spectral efficiency (SE) and provide massive connectivity in support of the escalating high data rate requirements imposed by  the unprecedented proliferation of mobile devices \cite{J. G. Andrews}. Cognitive radio (CR) and non-orthogonal multiple access (NOMA) constitute promising techniques of achieving high SE \cite{Y. C. liang}-\cite{Z. Ding}. Specifically, CR enables the secondary users (SUs) to exploit the frequency bands of the primary users (PUs) provided that the interference imposed on the PUs from the SUs is below a certain level.  NOMA has  a higher information-theoretic rate region than orthogonal techniques albeit, which is achieved  by increasing the receiver's implementation complexity \cite{Z. Ding}. One of the main ideas for realizing NOMA is to exploit the power domain. Specifically, multiple users' signals are superimposed by using different power levels and successive interference cancellation (SIC) is installed at the receiver for mitigating the mutual interference imposed by using non-orthogonal resources \cite{B. Wang}. It is envisioned that applying NOMA in CR networks (CRNs) is capable of significantly improving the SE and the user connectivity \cite{Y. Liu}, \cite{L. Lv}.

Meanwhile, the next generation wireless communication systems also need energy-efficient techniques due to the ever-increasing greenhouse gas emission concerns and explosive proliferation of power-limited devices, e.g., sensors and mobile phones. Energy-efficient techniques can be divided into two broad categories. One of the categories   focuses on the techniques that can achieve high energy efficiency (EE) \cite{F. H. Zhou}, \cite{D. Feng}, while the other one aims for recycling energy, where both wireless charging as well as simultaneous wireless information and power transfer (SWIPT) \cite{X. Lu} fit. In this paper, we focus on SWIPT since it can  simultaneously transmit information and achieve energy harvesting (EH).  In SWIPT,  the radio frequency (RF) signals carry not only  information to the users, but also transfer energy for the energy harvesting receivers (EHRs). Compared to the conventional EH techniques, such as wind charging, SWIPT has an advantage in providing more stable and controllable amount of power for energy-limited devices. Hence, it is of significant importance to study the application of SWIPT in NOMA CRNs that aim for supporting massive population of battery driven power-limited devices.

However, due to the broadcast nature of NOMA as well as CR and the dual function of RF signals \cite{Y. Wu}, \cite{Y. Zhang1}, NOMA CRNs relying on SWIPT are vulnerable to eavesdropping. Malicious EHRs may intercept the confidential information transmitted to the PUs and the SUs \cite{D. W. K. Ng1}. Thus, it is vital to improve the security of NOMA CRNs using SWIPT. As an alternative to the traditional cryptographic techniques, physical-layer security exploits the physical characteristics (e.g., multipath fading, propagation delay, etc.) of wireless channels to achieve secure communications \cite{X. Chen}-\cite{Y. Wu2}. It was shown \cite{X. Chen}-\cite{Y. Wu2} that the secrecy rate of wireless communication systems directly depends on the accuracy of the channel state information (CSI). Moreover, the secrecy rate of SUs in CRNs is more severely limited \cite{D. W. K. Ng1}, \cite{Y. Pei}-\cite{F. Zhou2} since their transmission power  should be controlled in order to protect the PUs' quality of service. In order to improve the secrecy rate of SUs, multiple antennas, cooperative relaying, jamming and artificial noise (AN)-aided techniques have been applied \cite{Y. Pei}-\cite{F. Zhou2}. Moreover, the secrecy rate can be further improved by designing an optimal resource allocation scheme \cite{Y. Pei}-\cite{F. Zhou2}. Furthermore, the secure energy efficiency can be enhanced by using AN-aided techniques and designing the optimal resource allocation schemes \cite{D. Wang}, \cite{D. Wang1}. However, the performance gains achieved by using these techniques are significantly influenced by the accuracy of CSI. What's worse, it is a challenge to obtain accurate CSI, especially for NOMA \cite{Q. Zhang}, \cite{F. Alavi}. Thus, it is important to design resource allocation schemes under the imperfect CSI.

Numerous investigations have been conducted for improving the security of the conventional OMA systems and efforts have been invested into conceiving secure NOMA systems \cite{Y. Zhang1}, \cite{B. He}-\cite{M. Tian}. However, no contributions have been  devoted to improving the security of NOMA CRNs using SWIPT. In this paper, in order to achieve secure communications, beamforming design problems are studied in multiple-input single-output (MISO) NOMA CRNs using SWIPT where a practical non-linear EH model is applied as well as different CSI models are considered. An AN-aided cooperative scheme is proposed for improving the security of the primary network. By using this scheme, the secondary network imposes artificial noise for jamming the malicious EHRs while aspiring to get a chance to access the frequency bands of the primary network. The related work and the motivation of our investigation are presented as follows.
\subsection{Related Work and Motivation}
Beamforming design problems have been extensively studied both in conventional CRNs \cite{Y. Pei1}-\cite{V. Nguyen} and in conventional CRNs using SWIPT \cite{D. W. K. Ng1}, \cite{F. Zhou2}, \cite{C. Xu}-\cite{Y. Huang}. Recently, some efforts have also been dedicated to designing NOMA resource allocation schemes for improving their security \cite{Y. Zhang1}, \cite{B. He}-\cite{M. Tian}. These contributions can be summarized as follows.

Due to the broadcast nature of the conventional CRNs, malicious SUs may intercept the confidential information  transmitted  to the legitimate SUs. In order to improve the security of CRNs, numerous secure physical-layer techniques have been proposed by using different CSI models \cite{Y. Pei1}-\cite{V. Nguyen}. In \cite{Y. Pei1}, a robust beamforming scheme has been proposed for MISO CRNs in the face of a bounded CSI error model. It was shown that as anticipated the secrecy rate of the SUs can be significantly improved by using multiple antennas techniques, by contrast it is reduced when the CSI inaccuracy goes up. By exploiting the relationship between multi-antenna aided secure communications and cognitive radio communications, the authors of \cite{L. Zhang} designed an optimal beamforming scheme for MISO-aided CRNs. In \cite{C. Wang}, the authors extended the contributions of \cite{Y. Pei1} and \cite{L. Zhang} into a fading channel and the secure throughput was maximized by optimizing both the beamforming vector and the transmission power. The authors of \cite{W. Zeng} studied the robust beamforming design problem in MISO CRNs where realistic finite-alphabet inputs are considered. A global optimization approach was proposed for designing an optimal beamforming scheme for maximizing the secrecy rate. Recently, the authors of \cite{D. W. K. Ng} and \cite{V. Nguyen} studied the beamforming design problems of secure MISO multiuser unicast CRNs and of mutlicast CRNs, respectively. Specifically, in \cite{D. W. K. Ng}, an AN-aided beamforming scheme was proposed. It was shown that as expected the secrecy rate of SUs can be improved by imposing artificial noise on malicious SUs. Cooperation between the primary network and the secondary network was proposed in \cite{V. Nguyen} where the secrecy rate of SUs was maximized  under the max-min fairness criterion.

Since energy harvesting has not been considered in \cite{Y. Pei1}-\cite{V. Nguyen}, the beamforming schemes proposed in these works are inappropriate in CRNs using SWIPT. Recently, the authors of \cite{D. W. K. Ng1}, \cite{F. Zhou2}, \cite{C. Xu}-\cite{Y. Huang}  studied the resource allocation problems of various CRNs using SWIPT. In \cite{D. W. K. Ng1}, a multi-objective optimization framework was applied in MISO CRNs with SWIPT. The beamforming scheme, the covariance matrix of AN and energy signals were jointly optimized. It was shown that there are several tradeoffs in CRNs using SWIPT, such as the tradeoff between the secrecy rate of SUs and the harvested power of EHRs. The authors of \cite{D. W. K. Ng1} only considered the bounded CSI error model. In \cite{F. Zhou2}, the authors studied the robust beamforming design problem both under the bounded CSI error model and the probabilistic CSI error model. It was shown that a performance gain can be obtained under the probabilistic CSI error model compared to the bounded CSI error model. Mohjazi \emph{et al.} \cite{L. Mohjazi} extended the robust beamforming design problem into a multi-user MISO CRNs using SWIPT. The transmission power of the cognitive base station (CBS) was minimized by jointly optimizing the beamforming of CBS and the power splitting factor of the energy-harvesting SUs. In order to further improve the secrecy rate and the harvested power of EHRs, an optimal precoding scheme was designed for multiple-input multiple-output (MIMO) aided CRNs using SWIPT \cite{C. Xu}. In \cite{H. Zhang}, a cooperative mechanism and a robust beamforming scheme were proposed for improving the security of CRNs, where the energy signals were exploited to jam the malicious EHRs. The authors of \cite{Y. Yuan} have studied robust resource allocation problems in MIMO-aided CRNs using SWIPT under the probabilistic CSI error model. The contributions of \cite{D. W. K. Ng1}, \cite{F. Zhou2}, \cite{C. Xu}-\cite{Y. Yuan} assumed an ideal linear EH model. However, practical power conversion circuits have a non-linear end-to-end wireless power transfer function. Hence, the robust resource allocation schemes proposed in these treatises would perform difficultly in the face of a realistic non-linear power transfer characteristic. In \cite{Y. Huang}, the robust beamforming design problem was studied in MISO CRNs using SWIPT,  where a non-linear EH model was used.

The above-mentioned contributions were made for CRNs and CRNs with SWIPT where OMA is applied. However, these resource allocation schemes proposed in the above-mentioned works are inappropriate or suboptimal in NOMA systems since NOMA schemes are very different from OMA. The authors of \cite{Y. Zhang1}, \cite{B. He}-\cite{M. Tian} have studied the optimal resource allocation problems in NOMA systems in order to achieve secure communications. In \cite{Y. Zhang1}, an optimal power allocation scheme was proposed for maximizing the secrecy sum rate of a single-input single-output (SISO) NOMA system, where only an eavesdropper was assumed and a constant decoding order was applied. In \cite{B. He}, the authors considered a more general scenario, where a dynamic decoding order was considered. The sum secrecy rate was maximized by jointly optimizing the decoding order, the transmission rates and the power allocated to users. The secrecy rate maximization problems of MISO NOMA systems \cite{Y. Li}, \cite{J. Lei} and MIMO NOMA systems \cite{M. Tian} were investigated. It was shown that the secrecy rate achieved by using NOMA is higher than that achieved by using OMA, and that the secrecy rate of users can be improved by using multiple antennas-aided techniques.

Although resource allocation problems have indeed been conceived for NOMA systems for achieving secure communications \cite{Y. Zhang1}, \cite{B. He}-\cite{M. Tian}, resource allocation schemes proposed in these contributions operated under the assumption that perfect CSI can be obtained. Moreover, these resource allocation schemes cannot work in NOMA CRNs using SWIPT since the interference between the primary network and the secondary network as well as the energy harvesting requirements of the EHRs are required to be considered. Furthermore, the robust resource allocation schemes proposed in conventional CRNs using SWIPT are inappropriate for NOMA CRNs using SWIPT due to the differences between NOMA and OMA. To the best of our knowledge, few investigations have been conduced for improving the security of NOMA CRNs using SWIPT. Thus, in order to achieve secure communications in NOMA CRNs using SWIPT, beamforming design problems are studied both under the perfect CSI model and the bounded CSI error model. These problems are challenging but meaningful. The reasons are from the following two perspectives. On the one hand, a practical non-linear EH model is applied, but the EH form is more complex than the linear form. On the other hand,  the mutual interference between the primary network and the secondary network as well as the interference among NOMA SUs have to be considered.

\subsection{Contributions and Organization}
In contrast to \cite{Y. Zhang1}, \cite{B. He}-\cite{M. Tian}, this paper studies the beamforming design problems of MISO-NOMA CRNs using SWIPT,  where multiple malicious EHRs exist and a practical non-linear EH model is applied. Both the perfect CSI and the bounded CSI error model are considered. In order to improve the security of the primary network, an AN-aided cooperative scheme is proposed. The main contributions are summarized as follows:

\begin{enumerate}
  \item The AN-aided cooperative scheme is proposed for MISO-NOMA CRNs using SWIPT in order to improve the security of the primary network. By using this scheme, the CBS transmits a jamming signal to cooperate with the primary base station (PBS) for improving the security of the PUs. As a reward, the secondary network is granted to access the frequency bands of the primary network and provide SWIPT services both for the SUs and for the EHRs in the secondary network. Moreover, the covariance matrix of the jamming signals transmitted at CBS and the beamforming of the CBS and the PBS are jointly optimized.
 \item Beamforming design problems are studied under both the perfect CSI model and the bounded CSI error model. In contrast to the works that only an eavesdropper was considered in the NOMA system \cite{Y. Zhang1}, \cite{B. He}-\cite{M. Tian}, we investigate a more general scenario, where multiple malicious EHRs exist. The total transmission power is minimized by jointly optimizing the transmission beamforming vectors of both the PBS and the CBS as well as the covariance matrix of the jamming signal transmitted at the CBS, subject to constraints on the secrecy rates of both the PUs and the SUs as well as on the energy harvesting requirements of the EHRs. A pair of algorithms are proposed for solving these challenging non-convex problems. One of them relies on semidefinite relaxation (SDR) while the other is based on a carefully conceived cost function.
 \item Our simulation results show that the proposed AN-aided cooperative scheme can reduce the transmission power required in MISO-NOMA CRNs using SWIPT. Moreover, it is shown that the performance achieved  by NOMA is proven to be better than that obtained by OMA, even when the CSI is imperfect. Furthermore, our simulation results also show that the algorithm based on the cost function outperforms the algorithm based on using SDR.
\end{enumerate}

\subsection{Organization and Notations}
The remainder of this paper is organized as follows. The system model is presented in Section II. Our secure beamforming design problems are examined under the perfect CSI assumption in Section III. Section IV presents our secure beamforming design problems under the bounded CSI error model while our simulation results are presented in Section V. Finally, the paper is concluded in Section VI.

\emph{Notations:} Vectors and matrices are represented by boldface lower case letters and boldface capital letters, respectively. The identity matrix is denoted by $\mathbf{I}$; $N_{P,t}$ and $N_{S,t}$ are the number of antennas of the PBS and the CBS, respectively; vec(\textbf{A}) denotes the vectorization of matrix \textbf{A} and it is obtained by stacking its column vectors. The Hermitian (conjugate) transpose, trace, and rank of a matrix \textbf{A} are denoted respectively by $\mathbf{A^H}$, Tr$\left(\mathbf{A}\right)$ and Rank$\left(\mathbf{A}\right)$. $\mathbf{x}^\dag$ represents the conjugate transpose of a vector $\mathbf{x}$. $\mathbf{C}^{M\times N}$ stands for an $M$-by-$N$ dimensional complex matrix set. $\mathbf{A}\succeq \mathbf{0} \left(\mathbf{A}\succ \mathbf{0}\right)$ represents that $\mathbf{A}$ is a Hermitian positive semidefinite (definite) matrix. $\mathbb{H}^N$ and $\mathbb{H}_+^{N}$ represent a $N$-by-$N$ dimensional Hermitian matrix set and a Hermitian positive semidefinite matrix set, respectively. ${\left\|  \cdot  \right\|}$ denotes the Euclidean norm of a vector. ${\left| \cdot \right|}$ represents the absolute value of a complex scalar. $\mathbf{x} \sim {\cal C}{\cal N}\left( {\mathbf{u},\mathbf{\Sigma } }\right)$ means that $\mathbf{x}$ is a random vector, which follows a complex Gaussian distribution with mean $\mathbf{u}$ and covariance matrix $\mathbf{\Sigma }$. $\mathbb{E}[ \cdot ]$ denotes the expectation operator. ${\rm Re}\left(\mathbf{a}\right)$ extracts the real part of vector $\mathbf{a}$. ${\lambda _{\max }}\left( {{\mathbf{A}}}\right) $ is the maximum eigenvalue of $\mathbf{A}$. $\mathbb{R}_{+}$ represents the set of all nonnegative real numbers. $[x]^{+}$ denotes the maximum between $0$ and $x$.

\section{System Model}
In this section, we will describe the network model and security metrics in the downlink MISO NOMA CRNs using SWIPT under a practical non-linear energy harvesting model. In \cite{Y. Zhang1}, \cite{B. He}-\cite{M. Tian}, only one eavesdropper has been considered in the designed NOMA systems and resource allocation schemes have been proposed. In this paper,  the beamforming design problems are studied in a more general scenario, where multiple malicious EHRs exist. The detail description is presented in the following subsections.
\subsection{Network Model}
\begin{figure}[!t]
\centering
\includegraphics[width=3.0 in]{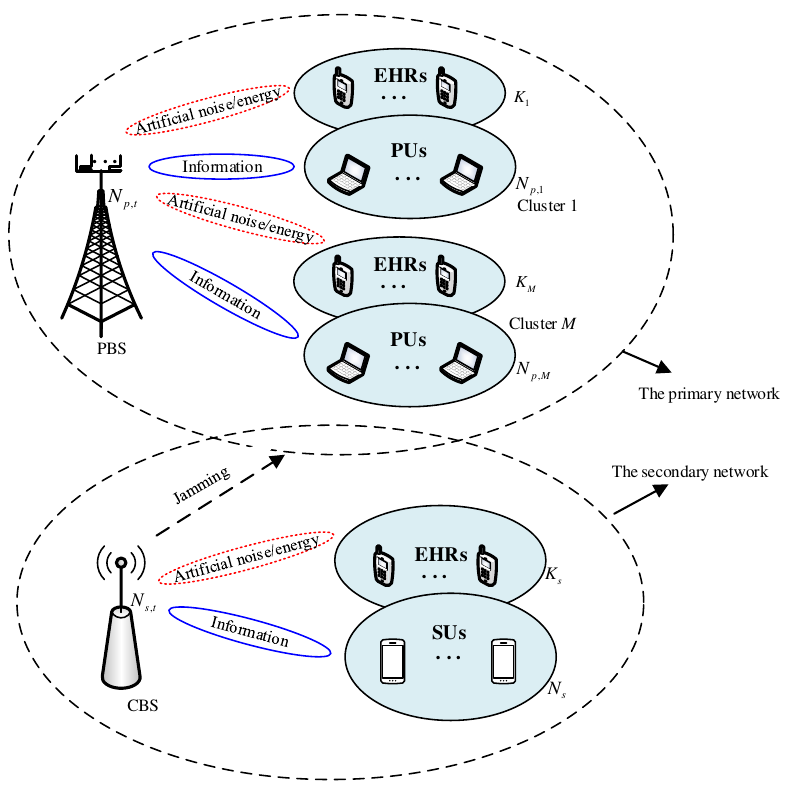}
\caption{The system model.} \label{fig.1}
\end{figure}
Our downlink MISO NOMA CR network using SWIPT is shown in Fig. 1. In the primary network, unicast-multicast communications are exploited since they can provide high SE and massive connectivity. This scenario is widely encountered, for example in Internet of Things, wireless sensor networks and the cellular networks \cite{D. W. K. Ng}, \cite{V. Nguyen}. Specifically, the PBS sends different  confidential
information-bearing signals to the PUs in the different clusters. And the primary users in each individual multicast cluster receive the same confidential information-bearing signal from the PBS. In the secondary network, the NOMA is applied since it can achieve high power transfer efficiency and SUs can perform SIC \cite{Y. Liu}, \cite{L. Lv}. In this case, the PBS broadcasts the information to the PUs in $M$ clusters and simultaneously transfers energy to EHRs. In the secondary network, the CBS provides SWIPT service to $K_s$ EHRs and to $N_s$ SUs by using NOMA. Due to the constrained size of devices, the PUs and SUs can only perform information decoding while the EHRs can only harvest energy from the RF signals \cite{F. Zhou2}, \cite{D. W. K. Ng}. The primary network coexists with the secondary network by using the spectrum sharing mode. The PBS is equipped  with $N_{p,t}$ antennas and the CBS is equipped with $N_{s,t}$ antennas. All the PUs, SUs and EHRs are equipped with a single antenna.

Due to the broadcast natures of NOMA and the dual function of RF signals in SWIPT, the EHR may eavesdrop and intercept the information transmitted by the PBS and the CBS. It is assumed that EHRs in each network can only intercept confidential information from the same network and the PUs in each cluster are respectively wiretapped by EHRs in the same cluster \cite{V. Nguyen}. For example, PUs in the $m$th cluster, where $m\in {\cal M}$ and $ {\cal M}\buildrel \Delta \over = \left\{ {1,2, \cdots ,M} \right\}$, are wiretapped by the $k$th EHR in the $m$th cluster, where $k\in {\cal M}$ and $ {\cal K}_m\buildrel \Delta \over = \left\{ {1,2, \cdots ,K_m} \right\}$ and $K_m$ is the number of EHRs while $N_{p,m}$ is the number of PUs in the $m$th cluster. In order to improve the security of both the primary network and the secondary network, an AN-aided cooperative scheme is applied. Using this scheme, the CBS of Fig. 1 transmits a jamming signal to the primary network for improving the security of the PUs. As a reward, the primary network allows the secondary network to operate on its frequency bands. All the channels involved are assumed to be flat fading channels. In this paper, both the perfect CSI and imperfect CSI cases are studied. The performance achieved under the perfect CSI can be used as a bound in our analysis and provides meaningful insights into the design of MISO NOMA CRNs using SWIPT. The  assumption has also been used in \cite{Y. Wu}, \cite{Y. Zhang1}, \cite{Y. Pei}, \cite{Y. Wu3}.

\subsection{Security Metrics}
Let ${y_{p,m,i}}$ denote the signal received at the $i$th PU in the $m$th cluster, ${y_{s,j}}$  represent the signal received at the $j$th SU, ${y_{{e},m,k}}$ denote the EH signal received at the $k$th EHR in the $m$th cluster and ${y_{e,l}}$ represent the EH signal received at the $l$th EHR in the secondary network, respectively, where $i\in {\cal N}_{p,m}$, ${\cal N}_{p,m}=\left\{ {1,2, \cdots ,N_{p,m}} \right\}$; $j\in {\cal N}_{s}$, ${\cal N}_{s}=\left\{ {1,2, \cdots ,N_{s}} \right\}$ and $l\in {\cal K}_{s}$, ${\cal K}_{s}=\left\{ {1,2, \cdots ,K_s} \right\}$. These signals are respectively expressed as
\begin{subequations}
\begin{align}\label{27}\ \notag
{y_{p,m,i}} =& \mathbf{h}_{p,m,i}^\dag \left[ {\sum\limits_{m = 1}^M \left({{\mathbf{w}_{p,m}}{s_{p,m}}}
 + {\mathbf{v}_{p,m}}\right)} \right]  \\
 & + \mathbf{f}_{s,m,i}^\dag \left( {\sum\limits_{j = 1}^{N_s} {{\mathbf{w}_{s,j}}{s_{s,j}}}  + {\mathbf{v}_s}} \right) + {n_{p,m,i}}, \\ \notag
{y_{s,j}} =& \mathbf{q}_{p,j}^\dag \left[ {\sum\limits_{m = 1}^M \left({{\mathbf{w}_{p,m}}{s_{p,m}}}  + {\mathbf{v}_{p,m}}\right)} \right]  \\
&+ \mathbf{h}_{s,j}^\dag \left( {\sum\limits_{j = 1}^{N_s} {{\mathbf{w}_{s,j}}{s_{s,j}}}  + {\mathbf{v}_s}} \right) + {n_{s,j}}, \\ \notag
\end{align}
\end{subequations}
\begin{subequations}
\begin{align}\label{27}\ \notag
{y_{e,m,k}} =& \mathbf{g}_{e,m,k}^\dag \left[ {\sum\limits_{m = 1}^M \left({{\mathbf{w}_{p,m}}{s_{p,m}}}  + {\mathbf{v}_{p,m}}\right)} \right] \\ \notag
&\tag{1c}+ \mathbf{f}_{e,m,k}^\dag \left( {\sum\limits_{j = 1}^{N_s} {{\mathbf{w}_{s,j}}{s_{s,j}}}  + {\mathbf{v}_s}} \right), \\ \notag
{y_{e,l}} = & \mathbf{q}_{e,l}^\dag \left[ {\sum\limits_{m = 1}^M \left({{\mathbf{w}_{p,m}}{s_{p,m}}}  + {\mathbf{v}_{p,m}}\right)} \right] \\ \notag
 &\tag{1d} + \mathbf{g}_{e,l}^\dag \left( {\sum\limits_{j = 1}^{N_s} {{\mathbf{w}_{s,j}}{s_{s,j}}}  + {\mathbf{v}_s}} \right),
\end{align}
\end{subequations}
where $\mathbf{h}_{p,m,i}\in {\mathbf{C}^{{N_{p,t}} \times 1}}$ and $\mathbf{f}_{s,m,i}\in {\mathbf{C}^{{N_{s,t}} \times 1}}$ are the channel vector between the PBS and the $i$th PU as well as that between the CBS and the $i$th PU in the $m$th cluster, respectively; $\mathbf{q}_{p,j}\in {\mathbf{C}^{{N_{p,t}} \times 1}}$ and $\mathbf{h}_{s,j}\in {\mathbf{C}^{{N_{s,t}} \times 1}}$ denote the channel vector between the PBS and the $j$th SU as well as that between the CBS and the $j$th SU, respectively. Furthermore, $\mathbf{g}_{e,m,k}\in {\mathbf{C}^{{N_{p,t}} \times 1}}$ and $\mathbf{f}_{e,m,k}\in {\mathbf{C}^{{N_{s,t}} \times 1}}$ are the channel vector between the PBS and the $k$th EHR and that between the CBS and the $j$th EHR in the $m$th cluster, respectively; $\mathbf{q}_{e,l}\in {\mathbf{C}^{{N_{p,t}} \times 1}}$ and $\mathbf{g}_{e,l}\in {\mathbf{C}^{{N_{s,t}} \times 1}}$ represent the channel vector between the PBS and the $l$th EHR and that between the CBS and the $l$th EHR in the secondary network, respectively. Still regarding to $\left(1\rm{a}\right)$, ${s_{p,m}} \in {\mathbf{C}^{{1} \times 1}}$ and $\mathbf{w}_{p,m}\in {\mathbf{C}^{N_{p,t} \times 1}}$ are the confidential information-bearing signal for the PUs in the $m$th cluster and the corresponding beamforming vector, respectively. Furthermore, ${s_{s,j}} \in {\mathbf{C}^{{1} \times 1}}$ and $\mathbf{w}_{s,j}\in {\mathbf{C}^{N_{s,t} \times 1}}$ represent the confidential information-bearing signal delivered for the $j$th SU and the corresponding beamforming vector, respectively. Additionally, ${\mathbf{v}_{p,m}}$ and ${\mathbf{v}_s}$ denote the noise vector artificially generated by the PBS and the CBS. It is assumed that $\mathbb{E}[ {{{\left| {s_{p,m}} \right|}^2}} ] = 1$ and $\mathbb{E}[ {{{\left| {s_{s,j}} \right|}^2}} ] = 1$. It is also assumed that $\mathbf{v}_{p,m} \sim {\cal C}{\cal N}\left( {0,\mathbf{\Sigma}_{p,m} } \right)$ and $\mathbf{v}_s \sim {\cal C}{\cal N}\left( {0,\mathbf{\Sigma}_s}  \right)$, where $\mathbf{\Sigma}_{p,m}$ and $\mathbf{\Sigma}_s$ are the AN covariance matrix. In $\left(1\right)$, ${n_{p,m,i}} \sim {\cal C}{\cal N}\left( {0,\sigma _{p,m,i}^2} \right)$ and ${n_{s,j}} \sim {\cal C}{\cal N}\left( {0,\sigma _{s,j}^2} \right)$ respectively denote the complex Gaussian noise at the $i$th PU in the $m$th cluster and the $l$th SU.

The secrecy rate of the $i$th PU in the $m$th cluster and the secrecy rate of the $j$th SU, denoted by $R_{p,m,i}$ and ${R_{s,j}}$, respectively, can be expressed as
\begin{subequations}
\begin{align}\label{27}\ \notag
&{R_{p,m,i}} = \left[\log \left( {\frac{{{\Gamma _{p,m,i}}}}{{{\Gamma _{p,m,i}} - \text{Tr}\left( {{\mathbf{W}_{p,m}}{\mathbf{H}_{p,m,i}}} \right)}}} \right) \right.\\  \notag
&\tag{2a}\ \ \left.  - \mathop {\max }\limits_{k \in {{\cal K}_m}} \log \left( {\frac{{{\Gamma _{e,m,k}} + \sigma _{e,m,k}^2}}{{{\Gamma _{e,m,k}} - \text{Tr}\left( {{\mathbf{W}_{p,m}}{\mathbf{G}_{e,m,k}}} \right) + \sigma _{e,m,k}^2}}} \right)\right]^{+},\\ \notag
&\tag{2b}{R_{s,j}} \left\{ \begin{array}{l}
=\left[{\log _2}\left( {\frac{{{\Gamma _{s,j}}}}{{{\Gamma _{s,j}} - \text{Tr}\left( {{\mathbf{W}_{s,j}}{\mathbf{H}_{s,j}}} \right)}}} \right)\right. \\ \ \ \ \left. - \mathop {\max }\limits_{l \in \cal L} {\log _2}\left( {\frac{{{\Lambda _{e,l,j}}}}{{{\Lambda _{e,l,j}} -\text{Tr}\left( {{\mathbf{W}_{s,j}}{\mathbf{G}_{e,l}}} \right)}}} \right)\right]^{+}, \text{if} \ j = {N_s}, \\
=\left[\mathop {\min }\limits_{z \in \left\{ {j,j + 1,{N_s}} \right\}} {\log _2}\left( {\frac{{{\Lambda _{s,j,z}}}}{{{\Lambda _{s,j,z}} - \text{Tr}\left( {{\mathbf{W}_{s,j}}{\mathbf{H}_{s,z}}} \right)}}} \right) \right.\\
\ \ \left. - \mathop {\max }\limits_{l \in \cal L} {\log _2}\left( {\frac{{{\Lambda _{s,l,j}}}}{{{\Lambda _{s,l,j}} - \text{Tr}\left( {{\mathbf{W}_{s,j}}{\mathbf{G}_{e,l}}} \right)}}} \right)\right]^{+},\text{otherwise}.
\end{array} \right.
\end{align}
\end{subequations}
where $\mathbf{W}_{p,m}=\mathbf{w}_{p,m}\mathbf{w}_{p,m}^\dag$; $\mathbf{W}_{s,j}=\mathbf{w}_{s,j}\mathbf{w}_{s,j}^\dag$; $\mathbf{H}_{p,m,i}=\mathbf{h}_{p,m,i}\mathbf{h}_{p,m,i}^\dag$; $\mathbf{F}_{s,m,i}=\mathbf{f}_{s,m,i}\mathbf{f}_{s,m,i}^\dag$; $\mathbf{Q}_{p,j}=\mathbf{q}_{p,j}\mathbf{q}_{p,j}^\dag$; $\mathbf{H}_{s,j}=\mathbf{h}_{s,j}\mathbf{h}_{s,j}^\dag$; $\mathbf{G}_{e,m,k}=\mathbf{g}_{e,m,k}\mathbf{g}_{e,m,k}^\dag$; $\mathbf{F}_{e,m,k}=\mathbf{f}_{e,m,k}\mathbf{f}_{e,m,k}^\dag$; $\mathbf{Q}_{e,l}=\mathbf{q}_{e,l}\mathbf{q}_{e,l}^\dag$ and $\mathbf{G}_{e,l}=\mathbf{g}_{e,l}\mathbf{g}_{e,l}^\dag$. The expressions of $\Gamma _{p,m,i}$, $\Gamma _{e,m,k}$, $\Gamma _{s,j}$, $\Lambda _{e,l,j}$, $\Lambda _{s,j,z}$ and $\Lambda _{s,l,j}$ are given in $\left(3\right)$. Without loss of generality, it is assumed that $\left\| {{\mathbf{h}_1}} \right\| \leq \left\| {{\mathbf{h}_2}} \right\|\leq\cdots \leq \left\| {{\mathbf{h}_{N_s}}} \right\|$. Similar to \cite{Y. Zhang1}, \cite{B. He}-\cite{Y. Li}, it is assumed furthermore that the EHR in the secondary network has decoded  SU $j$'s message before it decodes the SU  $i$'s  message, $j<i$. This over-estimates the interception capability of EHRs and results in the worst-case secrecy rate of the SUs. This conservative assumption was also used  in \cite{Y. Zhang1}, \cite{B. He}-\cite{Y. Li}.
\begin{subequations}
\begin{align}\label{27}\ \notag
&{\Gamma _{p,m,i}} = \text{Tr}\left\{ \left[ {\sum\limits_{m = 1}^M {\left( {{\mathbf{W}_{p,m}} + {\mathbf{\Sigma} _{p,m}}} \right)} } \right]{\mathbf{H}_{p,m,i}}\right.\\ \notag
 &\tag{3a} \ \ \ \ \ \ \ \ \ \ \left.+ \left( {\sum\limits_{j = 1}^{{N_s}} {{\mathbf{W}_{s,j}}}  + {\mathbf{\Sigma} _s}} \right){\mathbf{F}_{s,m,i}} \right\} + \sigma _{p,m,i}^2,\\ \notag
&{\Gamma _{e,m,k}} = \text{Tr}\left\{ \left[ {\sum\limits_{m = 1}^M {\left( {{\mathbf{W}_{p,m}} + {\mathbf{\Sigma} _{p,m}}} \right)} } \right]{\mathbf{G}_{e,m,k}}\right. \\ \notag
 &\tag{3b} \ \ \ \ \ \ \ \ \ \ \left.+ \left( {\sum\limits_{j = 1}^{N_s} {{\mathbf{W}_{s,j}}}  + {\mathbf{\Sigma} _s}} \right){\mathbf{F}_{e,m,k}} \right\},\\ \notag
&{\Gamma _{s,j}} = \text{Tr}\left\{ \left[ {\sum\limits_{m = 1}^M {\left( {{\mathbf{W}_{p,m}} + {\mathbf{\Sigma} _{p,m}}} \right)} } \right]{\mathbf{Q}_{p,j}}\right.\\ \notag
&\tag{3c}\ \ \ \ \ \ \  \ \left. + \left( {{\mathbf{W}_{s,j}} + {\mathbf{\Sigma} _s}} \right){\mathbf{H}_{s,j}} \right\} + \sigma _{s,j}^2,\\ \notag
&{\Lambda _{e,l,j}} = \text{Tr}\left[ \left[ {\sum\limits_{m = 1}^M {\left( {{\mathbf{W}_{p,m}} + {\mathbf{\Sigma} _{p,m}}} \right)} } \right]{\mathbf{Q}_{e,l}}\right.\\  \notag
&\tag{3d}\ \ \ \ \ \ \ \ \ \left. + \left( {{\mathbf{W}_{s,j}} + {\mathbf{\Sigma} _s}} \right){\mathbf{G}_{e,l}} \right] + \sigma _{e,l}^2,\\ \notag
&{\Lambda _{s,j,z}} = \text{Tr}\left\{ \left[ {\sum\limits_{m = 1}^M {\left( {{\mathbf{W}_{p,m}} + {\mathbf{\Sigma} _{p,m}}} \right)} } \right]{\mathbf{H}_{p,z}}\right. \\
&\tag{3e}\ \ \ \ \ \ \ \ \ \ \left.+ \left( {\sum\limits_{u = j}^{{N_s}} {{\mathbf{W}_{s,u}}}  + {\mathbf{\Sigma} _s}} \right){\mathbf{H}_{s,z}} \right\} + \sigma _{s,z}^2,\\ \notag
&{\Lambda _{s,l,j}} = \text{Tr}\left\{ \left[ {\sum\limits_{m = 1}^M {\left( {{\mathbf{W}_{p,m}} + {\mathbf{\Sigma} _{p,m}}} \right)} } \right]{\mathbf{Q}_{e,l}}\right. \\
 &\tag{3f}\ \ \ \ \ \ \ \ \ \ \left.+ \left( {\sum\limits_{\upsilon  = j}^{{N_s}} {{\mathbf{W}_{s,\upsilon }}}  + {\mathbf{\Sigma} _s}} \right){\mathbf{G}_{e,l}} \right\} + \sigma _{e,l}^2.
\end{align}
\end{subequations}

\subsection{Non-linear Energy Harvesting Model}
In this paper, a practical non-linear EH model is adopted. According to \cite{Y. Huang}-\cite{E. Boshkovska2}, the harvesting power of EHRs, denoted by ${\Phi _{E,{\rm A}}}$, can be formulated as:
\begin{subequations}
\begin{align}\label{27}\
&\tag{4a}{\Phi _{e,{\rm A}}} = \left( {\frac{{{\psi _{e,{\rm A}}} - P_{e,{\rm A}}^{\max }{\Psi _{e,{\rm A}}}}}{{1 - {\Psi _{e,{\rm A}}}}}} \right),\\
&\tag{4b}{\psi _{e,{\rm A}}} = \frac{{P_{e,{\rm A}}^{\max }}}{1 + {\exp{\left[{ - {a_{e,{\rm A}}}\left( {{\Gamma _{e,{\rm A}}} - {b_{e,{\rm A}}}} \right)}\right]}}},\\
&\tag{4c}{\Psi _{e,{\rm A}}} = \frac{1}{{1 + {\exp\left({{a_{e,{\rm A}}}{b_{e,{\rm A}}}}\right)}}},
\end{align}
\end{subequations}
where $A$ is the set of EHRs in the primary network and the secondary network, namely, $A = {A_1} \cup {A_2}$, and ${A_1} = \mathop  \cup \limits_{m \in \cal M} {\cal K}_m$, $m\in{\cal M}$, ${A_2} =  {\cal K}_{s}$; $a_{e,{\rm A}}$ and $b_{e,{\rm A}}$ represent parameters that reflect the circuit specifications, such as the resistance, the capacitance and diode turn-on voltage \cite{E. Boshkovska}. Furthermore, $P_{e,{\rm A}}^{\max }$ is the maximum harvested power of EHRs when the EH circuit is saturated. In $\left(4\rm{b}\right)$, ${\Gamma _{e,{\rm A}}} $ is the RF power received at EHRs. Furthermore, ${\Gamma _{e,{\rm A}}}= {\Gamma _{e,m,k}}$ when the EHRs are in the primary network and ${\Gamma _{e,{\rm A}}}= \Lambda _{s,l,1}-\sigma _{e,l}^2$ when the EHRs are in the secondary network. Note that the noise power is ignored, since it is small compared to the RF signal power \cite{Y. Huang}-\cite{E. Boshkovska2}.

\section{AN-aided Beamforming Design Under Perfect CSI}
In this section, an AN-aided beamforming design problem is formulated in MISO NOMA CRNs using SWIPT under the perfect CSI. The CSI between the PBS and PUs as well as the CSI between the CBS and the SUs can be obtained through the feedback from the corresponding transmitters and the receivers \cite{Y. Wu}, \cite{Y. Zhang1}, \cite{Y. Pei}, \cite{Y. Wu3}. The CSI between the two networks can be obtained with the cooperation between the primary network and the secondary network \cite{Y. Pei}, \cite{C. Wang}, \cite{W. Zeng}. The total transmission power is minimized subject to the constraints on both the secrecy rates of PUs and SUs as well as on the harvested power of EHRs in both the primary and the secondary networks. In order to solve the challenging non-convex problem, again,  a pair of suboptimal algorithms are proposed. One is based on SDR and the other is based on a cost function.
\subsection{AN-aided Beamforming Design Problem}
In order to minimize the sum of the transmission power of the PBS and CBS, the beamforming weights and the AN covariance of the PBS and the CBS are jointly optimized under constraints of  the secrecy rate of PUs as well as SUs and under the EH requirements of the EHRs. The power minimization problem is formulated as follows:
\begin{subequations}
\begin{align}\label{27}\
 &\tag{5a} \text{P}_{{1}}: \mathop {\min }\limits_{\scriptstyle{\mathbf{W}_{p,m}},{\mathbf{\Sigma} _{p,m}}\hfill\atop
\scriptstyle{\mathbf{W}_{s,j}},{\mathbf{\Sigma} _s}\hfill} \ {\text{Tr}\left[ {\sum\limits_{m = 1}^M {\left( {{\mathbf{W}_{p,m}} + {\mathbf{\Sigma} _{p,m}}} \right)}  + \sum\limits_{j = 1}^{N_s} {{\mathbf{W}_{s,j}}}  + {\mathbf{\Sigma} _s}} \right]}\\ \notag
&\text{s.t.}\ \\
&\tag{5b} C1:{R_{p,m,i}} \ge {\gamma _{p,m,i}},i \in {{\cal{N}}_{p,m}},m \in \cal{M},\\
&\tag{5c}  C2:{R_{s,j}} \ge {\gamma _{s,j}},j \in {{\cal{N}}_s},\\
&\tag{5d}  C3:{\Phi _{e,{A_1}}} \ge {\zeta _{e,{A_1}}},k \in {{{\cal{K}}{_{m}}}},m \in \cal{M},\\
&\tag{5e}  C4:{\Phi _{e,{A_2}}} \ge {\zeta _{e,{A_2}}},l \in {\cal{K}}_s,\\
&\tag{5f}  C5:\text{Rank}\left(\mathbf{W}_{p,m}\right)=1, \text{Rank}\left(\mathbf{W}_{s,j}\right)=1\\
&\tag{5g}  C6: \mathbf{W}_{p,m}\succeq \mathbf{0}, \mathbf{W}_{s,j}\succeq \mathbf{0}.
\end{align}
\end{subequations}
In $\left(5\right)$, ${\gamma _{p,m,i}}$ and $\gamma _{s,j}$ are the minimum secrecy rate requirements of the $i$th PU in the $m$th cluster and of the $j$th SU; $\zeta _{e,{A_1}}$ and $\zeta _{e,{A_2}}$ are the minimum EH requirements of EHRs in the primary and the secondary network. The constraints $C1$ and $C2$ are imposed to guarantee the secrecy rates of the PUs and SUs, respectively; the constraints $C3$ and $C4$ are the constraints that can satisfy the harvested power requirements of the EHRs in both the primary and secondary networks; and the constraint $C5$ is the rank-one constraint required for obtaining rank-one beamforming. Note that the optimization objective of $\text{P}_{{1}}$ can be identified as the weight objective of a multiple-objective optimization problem that has two optimization objectives (e.g., the transmission power of the PBS and the CBS)  with the same weight. Due to the constraints $C1$, $C2$ and $C5$, $\text{P}_{{1}}$ is non-convex and difficult to solve. In order to solve this problem, a pair of suboptimal schemes are proposed as follows.
\subsection{Suboptimal Solution Based on SDR}
To address the constraint $C1$, an auxiliary variable $\tau_m$, $m \in \cal{M}$, is introduced. Then, the constraint $C1$ can be equivalently
expressed as
\begin{subequations}
\begin{align}\label{27}\
&\tag{6a}\log \left\{ {\frac{{{\Gamma _{p,m,i}}}}{{\left[ {{\Gamma _{p,m,i}} - \text{Tr}\left( {{\mathbf{W}_{p,m}}{\mathbf{H}_{p,m,i}}} \right)} \right]{\tau _m}}}} \right\} \ge {\gamma _{p,m,i}},\\
&\tag{6b}\log \left\{ {\frac{{{\Gamma _{e,m,k}} + \sigma _{E,m,k}^2}}{{\left[ {{\Gamma _{e,m,k}} - \text{Tr}\left( {{\mathbf{W}_{p,m}}{\mathbf{G}_{e,m,k}}} \right) + \sigma _{e,m,k}^2} \right]{\tau _m}}}} \right\} \le 1,
\end{align}
\end{subequations}
where $k \in {{{\cal{K}}{_{m}}}}$ and $m \in \cal{M}$. Using successive convex approximation (SCA), the constraints given by $\left(6\rm{a}\right)$ and $\left(6\rm{b}\right)$ can be approximated as $\left(7\right)$ and $\left(8\right)$
\begin{subequations}
\begin{align}\label{27}\
&\tag{7a}\exp \left( {{\alpha _{p,m,i}} + {\beta _m} - {\lambda _{p,m,i}}} \right) \le {2^{ - {\gamma _{p,m,i}}}},\\ \notag
&{\Gamma _{p,m,i}} - \text{Tr}\left( {{\mathbf{W}_{p,m}}{\mathbf{H}_{p,m,i}}} \right) \\
&\tag{7b}\ \ \ \ \ \ \ \ \ \ \ \ \ \le\exp \left( {{{\widetilde \alpha }_{p,m,i}}} \right)\left( {{\alpha _{p,m,i}} - {{\widetilde \alpha }_{p,m,i}} + 1} \right),\\
&\tag{7c}{\tau _m} \le \exp \left( {{{\widetilde \beta }_m}} \right)\left( {{\beta _m} - {{\widetilde \beta }_m} + 1} \right),\\
&\tag{7d}{\Gamma _{p,m,i}} \ge \exp \left( {{\lambda _{p,m,i}}} \right),
\end{align}
\end{subequations}
\begin{subequations}
\begin{align}\label{27}\
&\tag{8a}\exp \left( {{\mu _{e,m,k}} - {\rho _{e,m,k}} - {\delta _m}} \right) \le 1,\\ \notag
&{\Gamma _{e,m,k}} + \sigma _{e,m,k}^2 \\
&\tag{8b}\ \ \ \ \ \ \ \ \ \ \ \ \ \le \exp \left( {{{\widetilde \mu }_{e,m,k}}} \right)\left( {{\mu _{e,m,k}} - {{\widetilde \mu }_{e,m,k}} + 1} \right),\\
&\tag{8c}{\Gamma _{e,m,k}} - \text{Tr}\left( {{\mathbf{W}_{p,m}}{\mathbf{G}_{e,m,k}}} \right) + \sigma _{e,m,k}^2 \ge \exp \left( {{\rho _{e,m,k}}} \right),\\
&\tag{8d}{\tau _m} \ge \exp \left( {{\delta _m}} \right),
\end{align}
\end{subequations}
where $\alpha _{p,m,i}$, ${\beta _m}$, $\lambda _{p,m,i}$, $\mu _{e,m,k}$, $\rho _{e,m,k}$,  and $\delta _m$ are auxiliary variables. Furthermore, ${\widetilde \alpha }_{p,m,i}$, ${\widetilde \beta }_m$ and ${\widetilde \mu }_{e,m,k}$ are approximate values, and they are equal to $\alpha _{p,m,i}$, ${\beta _m}$ and $\mu _{e,m,k}$, respectively, when the constraints are tight. Similarly, the constraint $C2$ can be approximated as $\left(9\right)$ and $\left(10\right)$. When $j=N_s$, the secrecy rate constraint of the $N_s$th SU can be formulated  as
\begin{subequations}
\begin{align}\label{27}\
&\tag{9a}\exp \left( {{\alpha _{s,{N_s}}} + {\beta _{s,{N_s}}} - {\lambda _{s,{N_s}}}} \right) \le {2^{ - {\gamma _{s,{N_s}}}}},\\ \notag
&{\Gamma _{s,{N_s}}} - \text{Tr}\left( {{\mathbf{W}_{s,{N_s}}}{\mathbf{H}_{S,{N_s}}}} \right) \\
 &\tag{9b}\ \ \ \ \ \ \ \ \ \ \le \exp \left( {{{\widetilde \alpha }_{s,{N_{s}}}}} \right)\left( {{\alpha _{s,{N_s}}} - {{\widetilde \alpha }_{s,{N_s}}} + 1} \right),\\
&\tag{9c}{\tau _{s,{N_s}}} \le \exp \left( {{{\widetilde \beta }_{s,{N_s}}}} \right)\left( {{\beta _{s,{N_s}}} - {{\widetilde \beta }_{s,{N_s}}} + 1} \right),\\
&\tag{9d}{\Gamma _{s,{N_s}}} \ge \exp \left( {{\lambda _{s,{N_s}}}} \right),\\
&\tag{9e}\exp \left( {{\mu _{e,l}} - {\rho _{s,l}} - {\omega _{s,{N_s}}}} \right) \le 1,l \in {\cal{K}}_s,\\
&\tag{9f}{\Lambda _{e,l,{N_s}}} \le \exp \left( {{{\widetilde \mu }_{e,l}}} \right)\left( {{\mu _{E,l}} - {{\widetilde \mu }_{e,l}} + 1} \right),\\
&\tag{9g}{\Lambda _{e,l,{N_s}}} - \text{Tr}\left( {{\mathbf{W}_{s,{N_s}}}{\mathbf{G}_{e,l}}} \right) \ge \exp \left( {{\rho _{s,l}}} \right),\\
&\tag{9h}{ \tau_{s,{N_s}}} \ge \exp \left( {{ \omega_{s,{N_s}}}} \right),
\end{align}
\end{subequations}
where $\alpha _{s,{N_s}}$, $\beta _{s,{N_s}}$, $\lambda _{s,{N_s}}$, $\mu _{e,l}$, $\rho _{s,l}$,  and $\omega _{s,{N_s}}$ are auxiliary variables. Furthermore, ${\widetilde \alpha }_{s,{N_s}}$, ${\widetilde \beta }_{s,{N_s}}$ and ${\widetilde \mu }_{e,l}$ are approximate values, and they are equal to $\alpha _{s,{N_s}}$, $\beta _{s,{N_s}}$ and $\mu _{e,l}$, respectively,  when the constraints are tight. When $j=1,2,\cdots,N_s-1$, the secrecy rate constraint of the $j$th SU can be formulated as
\begin{subequations}
\begin{align}\label{27}\
&\tag{10a}{\kappa _j} - {\omega _j}{2^{{\gamma _{s,j}}}} \ge 0,\\
&\tag{10b}\exp \left( {{\alpha _{s,j,z}} + {\xi _{s,j}} - {\lambda _{s,j,z}}} \right) \le 1,z \in \left\{ {j,j + 1,{N_s}} \right\},\\ \notag
&{\Lambda _{s,j,z}} - \text{Tr}\left( {{\mathbf{W}_{s,j}}{\mathbf{H}_{s,z}}} \right)\\
 &\tag{10c}\ \ \ \ \ \ \ \ \ \le \exp \left( {{{{\widetilde \alpha }_{s,j,z}}}} \right)\left( {{\alpha _{s,j,z}} - {{\widetilde \alpha }_{s,j,z}} + 1}\right),\\
&\tag{10d}{\kappa _j} \le \exp \left( {{{\widetilde \xi }_{s,j}}} \right)\left( {{\xi _{s,j}} - {{\widetilde \xi }_{s,j}} + 1} \right),\\
&\tag{10e}{\Lambda _{s,j,z}} \ge \exp \left( {{\lambda _{s,j,z}}} \right),\\
&\tag{10f}\exp \left( {{\mu _{e,l,j}} - {\rho _{s,l,j}} - {\tau _{s,j}}} \right) \le 1,\\
&\tag{10g}{\Lambda _{s,l,j}} \le \exp \left( {{{\widetilde \mu }_{e,l,j}}} \right)\left( {{\mu _{e,l,j}} - {{\widetilde \mu }_{E,l,j}} + 1} \right),\\
&\tag{10h}{\Lambda _{s,l,j}} - \text{Tr}\left( {{\mathbf{W}_{s,j}}{\mathbf{G}_{e,l}}} \right) \ge \exp \left( {{\rho _{s,l,j}}} \right),\\
&\tag{10i}{\omega _j} \ge \exp \left( {{\tau _{s,j}}} \right),
\end{align}
\end{subequations}
where $\kappa _j$, $\omega _j$, $\alpha _{s,j,z}$, $\xi _{s,j}$, $\lambda _{s,j,z}$, $\mu _{e,l,j}$, $\rho _{s,l,j}$  and $\tau _{s,j}$ denote auxiliary variables. Furthermore,  ${\widetilde \alpha }_{s,j,z}$, ${\widetilde \xi }_{s,j}$ and ${\widetilde \mu }_{e,l,j}$ are approximate values and equal to $\alpha _{s,j,z}$, $\xi _{s,j}$ and $\mu _{e,l,j}$, respectively, when the constraints are tight. Constraints $C3$ and $C4$ can be equivalently expressed as
\begin{align}\label{27}\
&\tag{11}{\Gamma _{e,A}}\ge {b_{e,A}} - \frac{1}{{{a_{e,A}}}}\ln \left\{ {\frac{{P_{e,A}^{\max }}}{{{\zeta _{e,A}}\left( {1 - {\Psi _{e,A}}} \right) + P_{e,A}^{\max }{\Psi _{e,A}}}} - 1} \right\}.
\end{align}
Based on $\left(7\right)$ and $\left(11\right)$, using SDR, $\text{P}_{{1}}$  can be solved by iteratively solving $\text{P}_{{2}}$, given as
\begin{subequations}
\begin{align}\label{27}\
 &\text{P}_{{2}}: \ \mathop {\min }\limits_{\Xi} \ {\text{Tr}\left[ {\sum\limits_{m = 1}^M {\left( {{\mathbf{W}_{p,m}} + {\mathbf{\Sigma} _{p,m}}} \right)}  + \sum\limits_{j = 1}^{N_s} {{\mathbf{W}_{s,j}}}  + {\mathbf{\Sigma} _s}} \right]}\\
&\text{s.t.}\ C6, \left(7\right)-\left(11\right),
\end{align}
\end{subequations}
where $\Xi$ is the set including all optimization variables and auxiliary variables.  $\text{P}_{{2}}$ is convex and can be efficiently solved by using the software \texttt{CVX} \cite{F. Zhou2}. Algorithm 1  calculates the solution of $\text{P}_{{1}}$. The details of Algorithm 1 are provided in Table 1, where $P_{opt}^n$ denotes the minimum total transmission power at the $n$th iteration.

\begin{table}[htbp]
\begin{center}
\caption{The SCA-based algorithm}
\begin{tabular}{lcl}
\\\toprule
$\textbf{Algorithm 1}$: The SCA-based algorithm for $\text{P}_{{1}}$\\ \midrule
\  1: \textbf{Setting:}\\
\ \  \ $\gamma _{p,m,i}$, $\gamma _{s,j}$ $\Upsilon _{m,i}$, $\zeta _{e,{A_1}}$, $\zeta _{e,{A_1}}$, $i \in {{\cal{N}}_{p,m}}$, $k \in {{{\cal{K}}{_{m}}}}$,  $m \in \cal{M}$\\
\ \  \ $l \in {\cal{K}}_s$ and the tolerance error $\varpi$; \\
\  2: \textbf{Initialization:}\\
\ \  \ The iterative number $n=1$, ${\widetilde \alpha }_{p,m,i}^n$, ${\widetilde \beta }_m^n$, ${\widetilde \mu }_{e,m,k}^n$, ${\widetilde \alpha }_{s,{N_s}}^n$, ${\widetilde \beta }_{s,{N_s}}^n$,\\
\ \  \ ${\widetilde \mu }_{e,l}^n$, ${\widetilde \alpha }_{s,j,z}^n$, ${\widetilde \xi }_{s,j}^n$ and ${\widetilde \mu }_{e,l,j}^n$ and $P_{opt}^n$; \\
\  3: \textbf{Repeat:}\\
 \   \ \ \ solve $\text{P}_{\textbf{2}}$ by using \texttt{CVX} for the given approximate values; \\
\ \ \ \ \ obtain ${\widetilde \alpha }_{p,m,i}^{n+1}$, ${\widetilde \beta }_m^{n+1}$, ${\widetilde \mu }_{e,m,k}^{n+1}$, ${\widetilde \alpha }_{s,{N_s}}^{n+1}$, ${\widetilde \beta }_{s,{N_s}}^{n+1}$, ${\widetilde \mu }_{e,l}^{n+1}$,\\
\ \  \ \ \ ${\widetilde \alpha }_{s,j,z}^{n+1}$, ${\widetilde \xi }_{s,j}^{n+1}$ and ${\widetilde \mu }_{e,l,j}^{n+1}$ and $P_{opt}^{n+1}$; \\
 \ \ \ \ \ if $\text{Rank}\left(\mathbf{W}_{p,m}\right)=1$ and $ \text{Rank}\left(\mathbf{W}_{s,j}\right)=1 $\\
  \ \ \ \ \ \ Obtain optimal $\mathbf{W}_{p,m}$ and $\mathbf{W}_{s,j}$;  \\
 \ \ \ \ \ else \\
 \ \ \ \ \ \ Obtain suboptimal $\mathbf{W}_{p,m}$ and $\mathbf{W}_{s,j}$;   \\
 \ \ \ \ \ end \\
 \ \ \ \ \ update the iterative number $n=n+1$;  \\
 \ \ \ \ \ calculate the total transmit power $P_{opt}^n$;  \\
\ \ \ \ \ if $\left|P_{opt}^n-P_{opt}^{\left(n-1\right)}\right|\leq \varpi$ \\
\ \ \ \ \  \  break;\\
\ \ \ \ \ end;\\
\  4: \textbf{Obtain resource allocation:}\\
 \ \ \ \ \ \ $\mathbf{W}_{p,m}$, $\mathbf{W}_{s,j}$, $\mathbf{\Sigma} _{p,m}$ and $\mathbf{\Sigma} _s$. \\
\bottomrule
\end{tabular}
\end{center}
\end{table}
Algorithm 1 does not guarantee that the optimal beamforming weights $\mathbf{w}_{p,m}$, $\mathbf{w}_{s,j}$ can be obtained. If $\mathbf{W}_{p,m}$ and $\mathbf{W}_{s,j}$ are of rank-one, the optimal beamforming scheme can be obtained by the eigenvalue decomposition and the obtained eigenvectors are optimal beamforming. If $\mathbf{W}_{p,m}$ and $\mathbf{W}_{s,j}$ are not of rank-one, the suboptimal beamforming vectors can be obtained by using the  Gaussian randomization procedure \cite{F. Zhou2}.
\subsection{Suboptimal Solution Based on Cost Function}
Since $\mathbf{W}_{p,m}$ and $\mathbf{W}_{s,j}$ are semi-positive definite, the ranks of $\mathbf{W}_{p,m}$ and $\mathbf{W}_{s,j}$ are equal to $1$ when their maximum eigenvalues are equal to its trace, namely, when we have $\text{Rank}\left(\mathbf{W}_{p,m}\right)=\lambda_{\max}\left(\mathbf{W}_{p,m}\right)$ and $\text{Rank}\left(\mathbf{W}_{s,j}\right)=\lambda_{\max}\left(\mathbf{W}_{s,j}\right)$; Otherwise, we have $\text{Rank}\left(\mathbf{W}_{p,m}\right)>\lambda_{\max}\left(\mathbf{W}_{p,m}\right)$ and $\text{Rank}\left(\mathbf{W}_{s,j}\right)>\lambda_{\max}\left(\mathbf{W}_{s,j}\right)$. Thus, the rank-one constraint can be equivalent to $\sum\limits_{m = 1}^M {\left[ {\text{Tr}\left( \mathbf{W}_{p,m} \right) - {\lambda _{\max }}\left( \mathbf{W}_{p,m} \right)} \right]} +\sum\limits_{j = 1}^{N_s} \left[ \text{Tr}\left( \mathbf{W}_{s,j} \right) - \lambda _{\max }\right. $ $\left.\left( \mathbf{W}_{s,j} \right) \right] \le 0$. From this insight, we can see that the smaller $\sum\limits_{m = 1}^M \left[ \text{Tr}\left( \mathbf{W}_{p,m} \right) - {\lambda _{\max }}\right.\\ \left.\left( \mathbf{W}_{p,m} \right) \right] +\sum\limits_{j = 1}^{N_s} \left[ \text{Tr}\left( \mathbf{W}_{s,j} \right)\right. \left.- {\lambda _{\max }}\left( \mathbf{W}_{s,j} \right) \right]$ is, the more likely that the rank-one constraints can be satisfied. By exploiting a cost function based approach, $\text{P}_{{2}}$ is reformulated into $\text{P}_{{3}}$  as
\begin{subequations}
\begin{align}\label{27}\ \notag
&\text{P}_{{3}}:  \mathop {\min }\limits_{\Xi} \ {\text{Tr}\left[ {\sum\limits_{m = 1}^M {\left( {{\mathbf{W}_{p,m}} + {\mathbf{\Sigma} _{p,m}}} \right)}  + \sum\limits_{j = 1}^{N_s} {{\mathbf{W}_{s,j}}}  + {\mathbf{\Sigma} _s}} \right]}\\ \notag
&\ \ \ \ \ \ \ \ \ \ + \ell\left\{ \sum\limits_{m = 1}^M {\left[ {\text{Tr}\left( \mathbf{W}_{p,m} \right) - {\lambda _{\max }}\left( \mathbf{W}_{p,m} \right)} \right]}\right.\\
&\ \ \ \ \ \ \ \ \ \ \ \ \  \left. +\sum\limits_{j = 1}^{N_s} {\left[ {\text{Tr}\left( \mathbf{W}_{s,j} \right) - {\lambda _{\max }}\left( \mathbf{W}_{s,j} \right)} \right]}\right\} \\
&\ \ \ \ \ \ \ \text{s.t.}\ \ \ \left(12\rm{b}\right),
\end{align}
\end{subequations}
where $\ell>0$ is a cost factor. It may be readily shown that the minimum value $\sum\limits_{m = 1}^M {\left[ {\text{Tr}\left( \mathbf{W}_{p,m} \right) - {\lambda _{\max }}\left( \mathbf{W}_{p,m} \right)} \right]}\\ +\sum\limits_{j = 1}^{N_s} {\left[ {\text{Tr}\left( \mathbf{W}_{s,j} \right) - {\lambda _{\max }}\left( \mathbf{W}_{s,j} \right)} \right]}$ can be obtained by using a large  $\ell$ value. Since $\lambda_{\max}\left(\mathbf{W}_{p,m} \right)$ and $\lambda_{\max}\left(\mathbf{W}_{s,j}\right)$  are convex, $\text{P}_{{3}}$ is non-convex. The following lemma is applied to solve the non-convex problem $\text{P}_{{3}}$.

\emph{\text{Lemma 1 }} \cite{S. P. Boyd}: Let ${\lambda _{\max }}\left( {{\mathbf{X}}}\right) $ and  ${\lambda _{\max }}\left( {{\mathbf{Y}}}\right) $ denote the maximum eigenvalue of $\mathbf{X}$ and $\mathbf{Y}$, respectively. If $\mathbf{X}$ and $\mathbf{Y}$ are semi-positive definite, then we have ${\lambda _{\max }}\left( \mathbf{X} \right) - {\lambda _{\max }}\left( \mathbf{Y} \right) \ge \mathbf{y}_{\max }^\dag \left( {\mathbf{X} - \mathbf{Y}} \right){\mathbf{y}_{\max }}$, where ${\mathbf{y}_{\max }}$ is the eigenvector related to the maximum eigenvalue of ${\mathbf{Y}}$.

Using Lemma 1, $\text{P}_{{3}}$ can be approximated as $\text{P}_{{4}}$.
\begin{subequations}
\begin{align}\label{27}\
&\text{P}_{{4}}:  {\mathop {\min }\limits_{\Xi} }\ {f\left( {\mathbf{W}_{p,m}^{n + 1}},{\mathbf{W}_{s,j}^{n + 1}} \right)} \\
&\ \ \ \ \ \ \ \text{s.t.}\ \ \ \left(12\rm{b}\right),
\end{align}
\end{subequations}
where $n$ is the iteration index and ${f\left( {\mathbf{W}_{p,m}^{n + 1}},{\mathbf{W}_{s,j}^{n + 1}} \right)}$ is given by
\begin{align}\label{27}\ \notag
&{f\left( {\mathbf{W}_{p,m}^{n + 1}},{\mathbf{W}_{s,j}^{n + 1}} \right)}\\ \notag
&= {\text{Tr}\left[ {\sum\limits_{m = 1}^M {\left( {{\mathbf{W}_{p,m}^{n + 1}} + {\mathbf{\Sigma} _{p,m}}} \right)}  + \sum\limits_{j = 1}^{N_s} {{\mathbf{W}_{s,j}^{n + 1}}}  + {\mathbf{\Sigma} _s}} \right]}\\ \notag
&\ \ \ + \ell\left\{ \sum\limits_{m = 1}^M \left[ \text{Tr}\left( {\mathbf{W}_{p,m}^{n + 1}} \right) - {\lambda _{\max }}\left( {\mathbf{W}_{p,m}^n} \right)\right. \right.\\  \notag
&\ \ \ \left. \left.- {{\left( {\mathbf{w}_{p,m}^n} \right)}^\dag }\left( {\mathbf{W}_{p,m}^{n + 1} - \mathbf{W}_{p,m}^n} \right)\mathbf{w}_{p,m}^n \right]\right.\\ \notag
&\  \ \  \left. + \sum\limits_{j = 1}^{N_s} \left[ \text{Tr}\left( {\mathbf{W}_{s,j}^{n + 1}} \right) - {\lambda _{\max }}\left( {\mathbf{W}_{s,j}^n} \right) \right.\right.\\
 &\ \ \ \left.\left. - {{\left( {\mathbf{w}_{s,j}^n} \right)}^\dag }\left( {\mathbf{W}_{s,j}^{n + 1} - \mathbf{W}_{s,j}^n} \right)\mathbf{w}_{s,j}^n \right] \right\},
\end{align}
where $\mathbf{w}_{p,m}^n$ and $\mathbf{w}_{s,j}^n$ are the eigenvectors related to the maximum eigenvalue of ${\mathbf{W}_{p,m}^n}$ and ${\mathbf{W}_{s,j}^n}$, respectively. It is seen that $\text{P}_{{4}}$ is convex and can be solved by using \texttt{CVX}. By solving $\text{P}_{{4}}$,  the iterative Algorithm 2 can be designed to solve $\text{P}_{{1}}$. The details of Algorithm 2 are presented in Table 2.
\begin{table}[htbp]
\begin{center}
\caption{The cost function-based algorithm}
\begin{tabular}{lcl}
\\\toprule
$\textbf{Algorithm 2}$: The cost function-based algorithm for $\text{P}_{{1}}$\\ \midrule
\  1: \textbf{Setting:}\\
\ \  \ $\gamma _{p,m,i}$, $\gamma _{s,j}$ $\Upsilon _{m,i}$, $\zeta _{e,{A_1}}$, $\zeta _{e,{A_1}}$, $i \in {{\cal{N}}_{p,m}}$, $k \in {{{\cal{K}}{_{m}}}}$,  $m \in \cal{M}$\\
\ \  \ $l \in {\cal{K}}_s$ and the tolerance error $\varpi$; \\
\  2: \textbf{Initialization:}\\
\ \  \ The iterative number $n=1$, ${\widetilde \alpha }_{p,m,i}^n$, ${\widetilde \beta }_m^n$, ${\widetilde \mu }_{e,m,k}^n$, ${\widetilde \alpha }_{s,{N_s}}^n$, ${\widetilde \beta }_{s,{N_s}}^n$, \\
\ \ \ $\mathbf{W}_{p,m}^{n }$ and $\mathbf{W}_{s,j}^{n }$; \\
\  3: \textbf{Repeat:}\\
 \   \ \ \ solve $\text{P}_{\textbf{4}}$ by using \texttt{CVX} for the given approximate values; \\
\ \ \ \ \ obtain ${\widetilde \alpha }_{p,m,i}^{n+1}$, ${\widetilde \beta }_m^{n+1}$, ${\widetilde \mu }_{e,m,k}^{n+1}$, ${\widetilde \alpha }_{s,{N_s}}^{n+1}$, ${\widetilde \beta }_{s,{N_s}}^{n+1}$, ${\widetilde \mu }_{e,l}^{n+1}$,\\
\ \ \ \ \ $\mathbf{W}_{p,m}^{\left(n+1\right) }$ and $\mathbf{W}_{s,j}^{\left(n+1\right) }$;  \\
  \ \ \ \ \ set $\ell=2\ell$;\\
 \ \ \ \ \ end if  \\
 \ \ \ \ \ update the iterative number $n=n+1$;  \\
 \ \ \ \ \ calculate the total transmit power $P_{opt}^n$;  \\
\ \ \ \ \ if $\text{Tr}\left( {\mathbf{W}_{p,m}^{n + 1}} \right) - {\lambda _{\max }}\left( {\mathbf{W}_{p,m}^n} \right)\leq \varpi$ \\
\ \ \ \ \ and $\text{Tr}\left( {\mathbf{W}_{s,j}^{n + 1}} \right) - {\lambda _{\max }}\left( {\mathbf{W}_{s,j}^n} \right)\leq \varpi$ \\
\ \ \ \ \  \  break;\\
\ \ \ \ \ end;\\
\  4: \textbf{Obtain resource allocation:}\\
 \ \ \ \ \ \  $\mathbf{w}_{p,m}^{n }$, $\mathbf{w}_{s,j}^{n }$, $\mathbf{\Sigma} _{p,m}$ and $\mathbf{\Sigma} _s$. \\
\bottomrule
\end{tabular}
\end{center}
\end{table}
\section{AN-aided Beamforming Design Under Imperfect CSI}
In this section, the AN-aided beamforming  design problem is studied in a more practical MISO NOMA CRN using SWIPT, where the CSI between the CBS and the EHRs in the primary network and the CSI between the PBS and the EHRs in the secondary network are imperfect due to the limited cooperation between the primary and the secondary network. Since the bounded CSI error can be readily applied to model the estimating errors \cite{D. W. K. Ng1}, \cite{Q. Zhang}, \cite{D. W. K. Ng}-\cite{C. Xu}, it is opted for this paper. Moreover, the CSI between the PBS and SUs and the CSI between the CBS and the PBS can be obtained through the cooperation between the primary network and the secondary network, or can be obtained from a third party such as a manager center \cite{F. H. Zhou}. Furthermore, the CSI of the secondary link can be obtained by estimating it at the CBS and then sending it back to the CBS through a feedback link, which assumed error-free in this simplified model \cite{F. Zhou2}, \cite{X. Kang}. Under the bounded error model, a robust AN-aided beamforming design problem is formulated. The non-convex problem is solved by using SCA and the ${\cal S}\text{-Procedure} $ \cite{S. P. Boyd}.
\subsection{Robust AN-aided Beamforming Design Problem Formulation}
Under the bounded error model, the channel vector ${\mathbf{q}_{e,l}}$ can be formulated as
\begin{subequations}
\begin{align}\label{27}\
&{\mathbf{q}_{e,l}} = { \mathbf{\overline q} _{e,l}} + \Delta {\mathbf{q}_{e,l}},l \in {{\cal{K}}_s},\\
&{\mathbf{\Psi} _{e,l}} \buildrel \Delta \over = \left\{ {\Delta {\mathbf{q}_{e,l}} \in {C^{{N_{p,t}} \times 1}}:\Delta \mathbf{q}_{e,l}^\dagger\Delta {\mathbf{q}_{e,l}} \le \pounds_{e,l}^2} \right\},
\end{align}
\end{subequations}
while the channel vector ${\mathbf{f}_{e,m,k}}$ can be expressed as
\begin{subequations}
\begin{align}\label{27}\
&{\mathbf{f}_{e,m,k}} = { \mathbf{\overline f }_{e,m,k}} + \Delta {\mathbf{f}_{e,m,k}},m \in M,k \in {{\cal{K}}_m},\\
&{\mathbf{\Psi} _{e,m,k}} \buildrel \Delta \over = \left\{ {\Delta {\mathbf{f}_{e,m,k}} \in {C^{{N_{s,t}} \times 1}}:\Delta \mathbf{f}_{e,m,k}^\dagger\Delta {\mathbf{f}_{e,m,k}} \le \pounds_{e,m,k}^2} \right\},
\end{align}
\end{subequations}
where ${ \mathbf{\overline q} _{e,l}}$ and ${ \mathbf{\overline f }_{e,m,k}}$ are the estimates of $\mathbf{q}_{e,l}$ and $\mathbf{f}_{e,m,k}$, respectively; ${\mathbf{\Psi} _{e,l}} $ and ${\mathbf{\Psi} _{e,m,k}} $ represent the uncertainty regions of the channel vectors $\mathbf{q}_{e,l}$ and $\mathbf{f}_{e,m,k}$, respectively; $\Delta {\mathbf{q}_{e,l}}$ and $\Delta {\mathbf{f}_{e,m,k}}$ denote the channel estimation errors of $\mathbf{q}_{e,l}$ and $\mathbf{f}_{e,m,k}$; $\pounds _{e,l}$ and $\pounds _{e,m,k}$ are the radii of the uncertainty regions ${\mathbf{\Psi} _{e,l}} $ and ${\mathbf{\Psi} _{e,m,k}} $, respectively.

Based on the bounded error models for ${\mathbf{q}_{e,l}}$ and ${\mathbf{f}_{e,m,k}}$, the power minimization problem subject to the constraints on the secrecy rates of the PUs and the SUs as well as on the harvested power requirements of EHRs in both the primary and secondary networks can be formulated as $\text{P}_{{5}}$, given as
\begin{subequations}
\begin{align}\label{27}\
 &\mathop {\min }\limits_{\scriptstyle{\mathbf{W}_{p,m}},{\mathbf{\Sigma} _{p,m}}\hfill\atop
\scriptstyle{\mathbf{W}_{s,j}},{\mathbf{\Sigma} _s}\hfill} \ {\text{Tr}\left[ {\sum\limits_{m = 1}^M {\left( {{\mathbf{W}_{p,m}} + {\mathbf{\Sigma} _{p,m}}} \right)}  + \sum\limits_{j = 1}^{N_s} {{\mathbf{W}_{s,j}}}  + {\mathbf{\Sigma} _s}} \right]}\\ \notag
&\text{s.t.}\ \\
&C6:{R_{p,m,i}} \ge {\gamma _{p,m,i}},i \in {{\cal{N}}_{p,m}},m \in {\cal{M}}, \forall {\Delta {\mathbf{f}_{e,m,k}}}\in {\mathbf{\Psi} _{e,m,k}}, \\
&  C7:{R_{s,j}} \ge {\gamma _{s,j}},j \in {{\cal{N}}_s}, \forall {\Delta{\mathbf{q}_{e,l}}}\in {\mathbf{\Psi} _{e,l}}, \\
&  C8:{\Phi _{e,{A_1}}} \ge {\zeta _{e,{A_1}}},k \in {{{\cal{K}}{_{m}}}},{m \in \cal{M}}, \forall {\Delta {\mathbf{f}_{e,m,k}}}\in {\mathbf{\Psi} _{e,m,k}},\\
&  C9:{\Phi _{e,{A_2}}} \ge {\zeta _{e,{A_2}}},l \in {\cal{K}}_s, \forall {\Delta{\mathbf{q}_{e,l}}}\in {\mathbf{\Psi} _{e,l}},\\
&  C10:\text{Rank}\left(\mathbf{W}_{p,m}\right)=1, \text{Rank}\left(\mathbf{W}_{s,j}\right)=1\\
&  C11: \mathbf{W}_{p,m}\succeq \mathbf{0}, \mathbf{W}_{s,j}\succeq \mathbf{0}.
\end{align}
\end{subequations}
The problem $\text{P}_{{5}}$ is more challenging to solve due to the infinite inequality constraints imposed by the uncertain regions, ${\mathbf{\Psi} _{e,l}}$ and ${\mathbf{\Psi} _{e,m,k}}$ and owing to the non-convex constraints $C6$- $C10$.
\subsection{Suboptimal Solution Based on Cost Function}
In order to make $\text{P}_{{5}}$ tractable, the ${\cal S}\text{-Procedure} $ of \cite{S. P. Boyd} is applied.

\emph{\text{Lemma 2 $\left({\cal S}\text{-Procedure} \right) $ }} \cite{S. P. Boyd}: Let $ {f_i}\left( \mathbf{z} \right) = {\mathbf{z}^\dag }{\mathbf{A}_i}\mathbf{z} + 2{\mathop{\mathbf{\rm Re}}\nolimits} \left\{ {\mathbf{b}_i^\dag \mathbf{z}} \right\} + {c_i}, i \in \left\{ {1,2} \right\}$, where $\mathbf{z}\in \mathbf{C}^{N\times 1}$, $\mathbf{A}_i \in \mathbb{H}^N$, $\mathbf{b}_i\in \mathbf{C}^{N\times 1}$ and $c_i\in \mathbb{R}$. Then, the expression ${f_1}\left( \mathbf{z} \right) \le 0 \Rightarrow {f_2}\left( \mathbf{z} \right) \le 0$ holds if and only if there exists a $\varsigma  \ge 0$ such that we have:
\begin{align}\label{27}\
\varsigma \left[ {\begin{array}{*{20}{c}}
{{\mathbf{A}_1}}&{{\mathbf{b}_1}}\\
{\mathbf{b}_1^\dag }&{{c_1}}
\end{array}} \right] - \left[ {\begin{array}{*{20}{c}}
{{\mathbf{A}_2}}&{{\mathbf{b}_2}}\\
{\mathbf{b}_2^\dag }&{{c_2}}
\end{array}} \right]
\succeq \mathbf{0},
\end{align}
provided that there exists a vector $\mathbf{\widehat z}$ so that we have ${f_i}\left( {\mathbf{\widehat z}} \right) < 0$.

Using Lemma 2 and SCA, the constraint $C6$ of   $\text{P}_{{5}}$ can be approximated as $\left(20\right)$ at the top of the next page.
\begin{figure*}[!t]
\normalsize
\begin{subequations}
\begin{align}\label{27}\
 &\left[ {\begin{array}{*{20}{c}}
{{{\overline \lambda  }_{e,m,k}}\mathbf{I} - \left( {\sum\limits_{j = 1}^{{N_s}} {{\mathbf{W}_{s,j}}}  + {\mathbf{\Sigma} _s}} \right)}&{ - \left( {\sum\limits_{j = 1}^{{N_s}} {{\mathbf{W}_{s,j}}}  + {\mathbf{\Sigma }_s}} \right){\mathbf{{\overline f }}_{e,m,k}}}\\
{ - \mathbf{\overline f} _{e,m,k}^{\dag}\left( {\sum\limits_{j = 1}^{{N_s}} {{\mathbf{W}_{s,j}}}  + {\mathbf{\Sigma} _s}} \right)}&{{\theta _{e,m,k}} - \mathbf{\overline f} _{e,m,k}^{\dag}\left( {\sum\limits_{j = 1}^{{N_s}} {{\mathbf{W}_{s,j}}}  + {\mathbf{\Sigma} _s}} \right){\mathbf{{\overline f }}_{e,m,k}} - {{\overline \lambda  }_{e,m,k}}\pounds_{e,m,k}^2}
\end{array}} \right]\succeq \mathbf{0 }, \\
&  \left[ {\begin{array}{*{20}{c}}
{{{\overline u }_{e,m,k}} \mathbf{I}+ \left( {\sum\limits_{j = 1}^{{N_s}} {{\mathbf{W}_{s,j}}}  + {\mathbf{\Sigma} _s}} \right)}&{\left( {\sum\limits_{j = 1}^{{N_s}} {{\mathbf{W}_{s,j}}}  + {\mathbf{\Sigma} _s}} \right){{\mathbf{\overline f }}_{e,m,k}}}\\
\mathbf{{{{\overline f}}}} _{e,m,k}^{\dag}\left( {\sum\limits_{j = 1}^{{N_s}} {{\mathbf{W}_{s,j}}}  + {\mathbf{\Sigma} _s}} \right)&{\mathbf{\overline f} _{e,m,k}^{\dag}\left( {\sum\limits_{j = 1}^{{N_s}} {{\mathbf{W}_{s,j}}}  + {\mathbf{\Sigma} _s}} \right){{\mathbf{\overline f }}_{e,m,k}} - {{\overline \lambda  }_{e,m,k}}\pounds_{e,m,k}^2 - {o_{E,m,k}}}
\end{array}} \right]\succeq \mathbf{0 },
\\
&  \ \ \ \ \ \ \ \ \ \ \ \ \ \ \ \ \ \ \ \ \ \ \ \ \ \ \ \ \ \ \ \ \ \ \ \ \ \ \ \ \ \ \ \ \ \ \ \ \ \ \ \ \ \ \ \ \ \ \ \ \ \ \ \ \ \ \ \ \ \ \ \ \ \ \ \ \ \ \ \ \ \ \ \ \left(7\right), \left(8\rm{a}\right)\ \text{and}\ \left(8\rm{d}\right).
\end{align}
\end{subequations}
\hrulefill \vspace*{4pt}
\end{figure*}
In $\left(20\right)$, ${{\overline \lambda  }_{e,m,k}}\geq0$ and ${\overline u }_{e,m,k}\geq0$ are slack variables while $\theta _{e,m,k}$ and ${o_{e,m,k}}$ are  auxiliary variables. Similarly, the constraint $C7$ is approximated as follows.

When $j=N_s$, one has $\left(21\right)$ at the top of the next page.
\begin{figure*}[!t]
\normalsize
\begin{subequations}
\begin{align}\label{27}\
 &\left[ {\begin{array}{*{20}{c}}
{{{\overline \omega  }_{e,l}}\mathbf{I} - \sum\limits_{m = 1}^M {\left( {{\mathbf{W}_{p,m}} + {\mathbf{\Sigma }_{p,m}}} \right)} }&{ - \sum\limits_{m = 1}^M {\left( {{\mathbf{W}_{p,m}} + {\mathbf{\Sigma} _{p,m}}} \right)} {{\mathbf{\overline q} }_{e,l}}}\\
{ - \mathbf{\overline q }_{e,l}^\dag \sum\limits_{m = 1}^M {\left( {{\mathbf{W}_{p,m}} + {\mathbf{\Sigma} _{p,m}}} \right)} }&{ - {{\overline \Lambda  }_{e,l}} - \mathbf{\overline q} _{e,l}^\dag \sum\limits_{m = 1}^M {\left( {{\mathbf{W}_{p,m}} + {\mathbf{\Sigma} _{p,m}}} \right){{\mathbf{\overline q} }_{e,l}} - {{\overline \omega  }_{e,l}}\pounds_{e,l}^2} }
\end{array}} \right]\succeq \mathbf{0 }, \\
&  \left[ {\begin{array}{*{20}{c}}
{{{\overline \kappa  }_{e,l}}\mathbf{I} + \sum\limits_{m = 1}^M {\left( {{\mathbf{W}_{p,m}} + {\mathbf{\Sigma} _{p,m}}} \right)} }&{\sum\limits_{m = 1}^M {\left( {{\mathbf{W}_{p,m}} + {\mathbf{\Sigma} _{p,m}}} \right)} {{\mathbf{\overline q }}_{e,l}}}\\
{\mathbf{\overline q} _{e,l}^\dag \sum\limits_{m = 1}^M {\left( {{\mathbf{W}_{p,m}} + {\mathbf{\Sigma} _{p,m}}} \right)} }&{{{\widetilde \Lambda }_{e,l}} + \mathbf{\overline q }_{e,l}^\dag \sum\limits_{m = 1}^M {\left( {{\mathbf{W}_{p,m}} + {\mathbf{\Sigma} _{p,m}}} \right){{\mathbf{\overline q} }_{e,l}} - {{\overline \kappa  }_{e,l}}\pounds_{e,l}^2} }
\end{array}} \right]\succeq \mathbf{0 }, \\
&{\overline \Lambda  _{e,l}} = \text{Tr}\left[ {\left( {{\mathbf{W}_{s,{N_{s}}}} + {\mathbf{\Sigma} _s}} \right){\mathbf{G}_{e,l}}} \right] + \sigma _{e,l}^2 - \exp \left( {{{\widetilde \mu }_{e,l}}} \right)\left( {{\mu _{e,l}} - {{\widetilde \mu }_{e,l}} + 1} \right), \\
&{\widetilde \Lambda _{e,l}} = \text{Tr}\left( {{\mathbf{\Sigma} _s}{\mathbf{G}_{e,l}}} \right) + \sigma _{e,l}^2 - \exp \left( {{\rho _{s,l}}} \right), \\
& \left(9\rm{a}\right)- \left(9\rm{d}\right), \left(9\rm{e}\right)\ \text{and}\ \left(9\rm{h}\right),
\end{align}
\end{subequations}
\hrulefill \vspace*{4pt}
\end{figure*}
and when $j=1,2,\cdots,N_s-1$, the secrecy rate constraint of the $j$th SU can be approximated as $\left(22\right)$ at the top of the next page.
\begin{figure*}[!t]
\normalsize
\begin{subequations}
\begin{align}\label{27}\
 &\left[ {\begin{array}{*{20}{c}}
{{{\overline \tau  }_{e,l}}\mathbf{I} - \sum\limits_{m = 1}^M {\left( {{\mathbf{W}_{p,m}} + {\mathbf{\Sigma} _{p,m}}} \right)} }&{ - \sum\limits_{m = 1}^M {\left( {{\mathbf{W}_{p,m}} + {\mathbf{\Sigma} _{p,m}}} \right)} {{\mathbf{\overline q} }_{e,l}}}\\
{ - \mathbf{\overline q} _{e,l}^\dag \sum\limits_{m = 1}^M {\left( {{\mathbf{W}_{p,m}} + {\mathbf{\Sigma} _{p,m}}} \right)} }&{ - {{\overline \Lambda  }_{s,l,j}} - \mathbf{\overline q} _{e,l}^\dag \sum\limits_{m = 1}^M {\left( {{\mathbf{W}_{p,m}} + {\mathbf{\Sigma} _{p,m}}} \right){\mathbf{{\overline q }}_{e,l}} - {{\overline \tau  }_{e,l}}\pounds_{e,l}^2} }
\end{array}} \right]
\succeq \mathbf{0 }, \\
&  \left[ {\begin{array}{*{20}{c}}
{{{\overline \eta  }_{e,l}}\mathbf{I} + \sum\limits_{m = 1}^M {\left( {{\mathbf{W}_{p,m}} + {\mathbf{\Sigma} _{p,m}}} \right)} }&{\sum\limits_{m = 1}^M {\left( {{\mathbf{W}_{p,m}} + {\mathbf{\Sigma} _{p,m}}} \right)} {{\mathbf{\overline q} }_{e,l}}}\\
{\mathbf{\overline q} _{e,l}^\dag \sum\limits_{m = 1}^M {\left( {{\mathbf{W}_{p,m}} + {\mathbf{\Sigma} _{p,m}}} \right)} }&{{{\overline \Lambda  }_{s,l,j}} + \mathbf{\overline q} _{e,l}^\dag \sum\limits_{m = 1}^M {\left( {{\mathbf{W}_{p,m}} + {\mathbf{\Sigma} _{p,m}}} \right){{\mathbf{\overline q} }_{e,l}} - {{\overline \eta  }_{e,l}}\pounds_{e,l}^2} }
\end{array}} \right]
\succeq \mathbf{0 }, \\
& {\overline \Lambda  _{s,l,j}} = \text{ Tr}\left\{ {\left( {\sum\limits_{\upsilon  = j}^{{N_s}} {{\mathbf{W}_{s,\upsilon }}}  + {\mathbf{\Sigma} _s}} \right){\mathbf{G}_{e,l}}} \right\} + \sigma _{e,l}^2 - \exp \left( {{{\widetilde \mu }_{e,l,j}}} \right)\left( {{\mu _{e,l,j}} - {{\widetilde \mu }_{e,l,j}} + 1} \right), \\
&{\widetilde \Lambda _{s,l,j}} = \text{ Tr}\left[ {\left( {\sum\limits_{\upsilon  = j + 1}^{{N_s}} {{\mathbf{W}_{s,\upsilon }}}  + {\mathbf{\Sigma} _s}} \right){\mathbf{G}_{e,l}}} \right] + \sigma _{e,l}^2 - \exp \left( {{\rho _{s,l,j}}} \right), \\
& \left(10\rm{a}\right)- \left(10\rm{f}\right), \ \text{and}\ \left(10\rm{i}\right),
\end{align}
\end{subequations}
\hrulefill \vspace*{4pt}
\end{figure*}
where ${{\overline \omega  }_{e,l}}\geq0$, ${\overline \kappa  }_{e,l}\geq0$, ${\overline \tau  }\geq0$ and ${\overline \eta  }_{e,l} \geq0$ are slack variables. The constraints $C8$ and $C9$ can be equivalently expressed as $\left(23\right)$ at the top of the next two pages.
\begin{figure*}[!t]
\normalsize
\begin{subequations}
\begin{align}\label{27}\
 &\left[ {\begin{array}{*{20}{c}}
{{{\overline \chi  }_{e,m,k}}\mathbf{I} + \left( {\sum\limits_{j = 1}^{{N_s}} {{\mathbf{W}_{s,j}}}  + {\mathbf{\Sigma} _s}} \right)}&{\left( {\sum\limits_{j = 1}^{{N_s}} {{\mathbf{W}_{s,j}}}  + {\mathbf{\Sigma} _s}} \right){{\mathbf{\overline f} }_{e,m,k}}}\\
{\mathbf{\overline f} _{e,m,k}^\dag\left( {\sum\limits_{j = 1}^{{N_s}} {{\mathbf{W}_{s,j}}}  + {\mathbf{\Sigma} _s}} \right)}&{{{\overline \Gamma  }_{e,m,k}} +\mathbf{ \overline f} _{e,m,k}^\dag\left( {\sum\limits_{j = 1}^{{N_s}} {{\mathbf{W}_{s,j}}}  + {\mathbf{\Sigma} _s}} \right){{\mathbf{\overline f }}_{E,m,k}} - {{\overline \chi  }_{e,m,k}}\pounds_{e,m,k}^2}
\end{array}} \right] \succeq \mathbf{0 }, \\
&  \left[ {\begin{array}{*{20}{c}}
{{{\overline \varphi  }_{e,l}}\mathbf{I} + \sum\limits_{m = 1}^M {\left( {{\mathbf{W}_{p,m}} + {\mathbf{\Sigma} _{p,m}}} \right)} }&{\sum\limits_{m = 1}^M {\left( {{\mathbf{W}_{p,m}} + {\mathbf{\Sigma} _{p,m}}} \right)} {{\mathbf{\overline q }}_{e,l}}}\\
{\mathbf{\overline q }_{e,l}^\dag \sum\limits_{m = 1}^M {\left( {{\mathbf{W}_{p,m}} + {\mathbf{\Sigma} _{p,m}}} \right)} }&{{{\overline \Gamma  }_{e,l}} +\mathbf{ \overline q} _{e,l}^\dag \sum\limits_{m = 1}^M {\left( {{\mathbf{W}_{p,m}} + {\mathbf{\Sigma} _{p,m}}} \right){{\mathbf{\overline q} }_{e,l}} - {{\overline \phi  }_{e,l}}\pounds_{E,l}^2} }
\end{array}} \right]\succeq \mathbf{0 }, \\ \notag
&{\overline \Gamma  _{e,m,k}} = \text{ Tr}\left\{ {\left[ {\sum\limits_{m = 1}^M {\left( {{\mathbf{W}_{p,m}} + {\mathbf{\Sigma} _{p,m}}} \right)} } \right]{\mathbf{G}_{e,m,k}}} \right\} - {b_{e,m,k}}m \\
 &\ \ \ \ \ \ \ \ \ \ \ \ \ + \frac{1}{{{a_{e,m,k}}}}\ln \left\{ {\frac{{P_{e,m,k}^{\max }}}{{{\psi _{e,m,k}}\left( {1 - {\Psi _{e,m,k}}} \right) + P_{e,m,k}^{\max }{\Psi _{E,m,k}}}} - 1} \right\}, \\
& {\overline \Gamma  _{e,l}} =\text{ Tr}\left\{ {\left( {\sum\limits_{j = 1}^{N_s} {{\mathbf{W}_{s,j}}}  + {\mathbf{\Sigma} _s}} \right){\mathbf{G}_{E,l}}} \right\} - {b_{e,l}}{\rm{ + }}\frac{1}{{{a_{e,l}}}}\ln \left\{ {\frac{{P_{e,l}^{\max }}}{{{\psi _{e,l}}\left( {1 - {\Psi _{e,l}}} \right) + P_{e,l}^{\max }{\Psi _{e,l}}}} - 1} \right\}.
\end{align}
\end{subequations}
\hrulefill \vspace*{4pt}
\end{figure*}
In $\left(23\right)$, ${{\overline \chi  }_{e,m,k}}\geq0$ and ${\overline \varphi  }_{e,l} \geq0$ are slack variables. By using $\left(20\right)$-$\left(23\right)$, $\text{P}_{{5}}$ can be solved by iteratively solving $\text{P}_{{6}}$, given as
\begin{subequations}
\begin{align}\label{27}\
 &\text{P}_{{6}}:{\mathop {\min }\limits_{\Xi_1} }\ {f\left( {\mathbf{W}_{p,m}^{n + 1}},{\mathbf{W}_{s,j}^{n + 1}} \right)}\\
\text{s.t.}\ & C11, \left(20\right)- \left(23\right),
\end{align}
\end{subequations}
where ${f\left( {\mathbf{W}_{p,m}^{n + 1}},{\mathbf{W}_{s,j}^{n + 1}} \right)}$ is given by $\left(15\right)$, and $\Xi_1$ denotes the set including all optimization variables, auxiliary variables and slack variables. Since $\text{P}_{{6}}$ is convex, it can be readily solved by using \texttt{CVX}. Similar to $\text{P}_{{1}}$, Algorithm 2 can be used to solve $\text{P}_{{5}}$. The procedure is the same and it is omitted due to space limitation.

\section{Simulation Results}
In this section, simulation results are provided for comparing the performance obtained by using NOMA to that achieved by using OMA. Simulation results are also presented to evaluate the performance of the proposed algorithms. The time division multiple access (TDMA) scheme is selected as the OMA scheme for comparison. The simulation settings are based on those used  in \cite{E. Boshkovska} and \cite{E. Boshkovska2}. All the channels involved are assumed to be Rayleigh flat fading. The number of channel realizations is $10^4$. The variance of noise at all users and EHRs is $-120$ dBm.  The channel distributions are set as: $\mathbf{h}_{p,m,i}\sim{\cal C}{\cal N}\left( {\mathbf{0},2\mathbf{I } }\right)$, $\mathbf{f}_{s,m,i} \sim{\cal C}{\cal N}\left( {\mathbf{0},0.5\mathbf{I } }\right)$, $\mathbf{q}_{p,j} \sim{\cal C}{\cal N}\left( {\mathbf{0},0.5\mathbf{I } }\right)$, $\mathbf{h}_{s,j} \sim{\cal C}{\cal N}\left( {\mathbf{0},2\mathbf{I } }\right)$, $\mathbf{g}_{e,m,k} \sim{\cal C}{\cal N}\left( {\mathbf{0},1.5\mathbf{I } }\right)$, $\mathbf{f}_{e,m,k} \sim{\cal C}{\cal N}\left( {\mathbf{0},0.5\mathbf{I } }\right)$, $\mathbf{q}_{e,l}\sim{\cal C}{\cal N}\left( {\mathbf{0},0.5\mathbf{I } }\right)$ and $\mathbf{g}_{e,l} \sim{\cal C}{\cal N}\left( {\mathbf{0},1.5\mathbf{I } }\right)$. The detailed simulation settings are given in Table III.

\begin{table}[htbp]
\centering
 \caption{\label{tab:test}Simulation Parameters}
 \begin{tabular}{l|c|c}
  \midrule
  \midrule
  Parameters & Notation & Typical Values  \\
  \midrule
  \midrule
 Numbers of antennas of the PBS & $N_{p,t}$ & $10$ \\
 Numbers of antennas of the CBS & $N_{s,t}$ & $5$ \\
 Numbers of the clusters & $M$ & $2$ \\
Numbers of PUs & $N_{p,m}$ & $2$ \\
 Numbers of SUs & $N_s$ & $3$ \\
 The maximum harvested power & $P_{e,A}^{\max }$ & $24$ mW \\
 Circuit parameter & $a_{e,A}$ & $1500$ \\
 Circuit parameter & $b_{e,A}$ & $0.0022$ \\
  The minimum secrecy rate of PUs & $\gamma _{p,m,i}$ & $2$ bits/s/Hz \\
 The minimum secrecy rate of SUs& $\gamma _{s,j}$ & $1$ bits/s/Hz \\
 The minimum EH of EHRs in set $A_1$& $\zeta _{e,A_1}$ & $15$ mW\\
 The minimum EH of EHRs in set $A_2$& $\zeta _{e,A_2}$ & $5$ mW\\
 The tolerance error & $\varpi$ & $10^{-4}$ \\
 \midrule
  The radiuses of the uncertainty regions& $\pounds_{e,l}$ & $10^{-2}$ \\
 &  $\pounds_{e,m,k}$ & $10^{-2}$ \\
\midrule
\midrule
 \end{tabular}
\end{table}

\begin{figure}[!t]
\centering
\includegraphics[width=3.5 in]{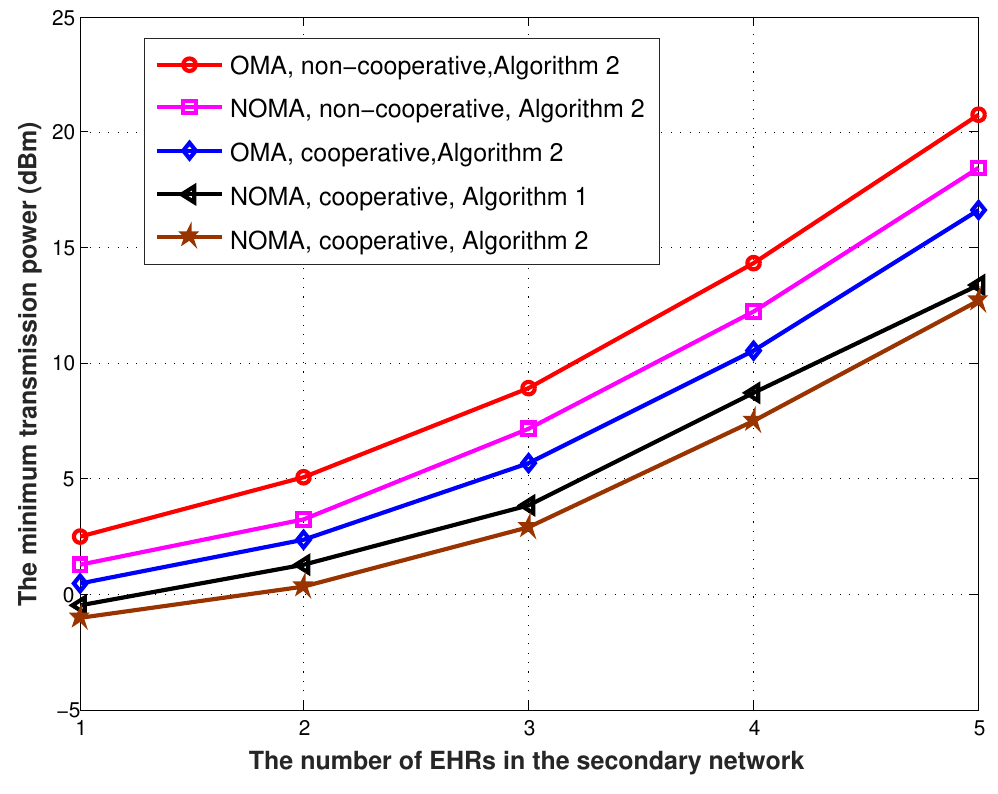}
\caption{The minimum transmission power versus the number of EHRs in the secondary network under the perfect CSI scenario.} \label{fig.1}
\end{figure}
Fig. 2 shows the minimum transmission power versus the number of EHRs in the secondary network under the perfect CSI scenario. Since Algorithm 2 is capable of obtaining rank-one solutions, it is used to obtain the results of CRNs using OMA, OMA with cooperation and NOMA without cooperation. Fig. 2 also compares the performance obtained by using Algorithm 1 to that achieved by using Algorithm 2. It is seen that the minimum transmission power consumed without the cooperative jamming scheme is higher than that consumed with our proposed cooperative jamming scheme. The reason is that our proposed cooperative jamming scheme is beneficial for the PUs to achieve a high secrecy rate and that a low transmission power is required for guaranteeing the secrecy rate of the PUs. This indicates that our proposed cooperative jamming scheme is eventually helpful for secure communications. It is also seen that the transmission power consumed by using NOMA is lower than that consumed by using TDMA both with and without cooperation between the primary network and the secondary network. This can be explained by the fact that NOMA provides a higher secrecy rate for SUs than TDMA \cite{Y. Zhang1}, \cite{B. He}. Thus, the transmission power required for guaranteeing the secrecy rate of SUs can be decreased. Moreover, it is interesting to note that the transmission power achieved by using Algorithm 1 is higher than that obtained by using Algorithm 2. It indicates that Algorithm 2 outperforms Algorithm 1 in terms of transmission power minimization. The reason is that Algorithm 2 can obtain rank-one solutions, while Algorithm 1 cannot achieve rank-one solutions.

\begin{table}
\begin{center}
\caption{Comparison of the number of rank-one solutions achieved by Algorithms 1 and 2}
\begin{tabular}{|c|c|c|c|c|c|}\hline
\backslashbox{Algorithm}{$\left(K_s\right)$} &$\left(1\right)$ &$\left(2\right)$ &$\left(3\right)$ &$\left(4\right)$&$\left(5\right)$\\\hline
Algorithm 1 &0 &0&0&0 &0 \\\hline
Algorithm 2&259 &207 &248&301&196 \\\hline
\end{tabular}
\end{center}
\end{table}
Table 3 is given to show the number of rank-one solutions achieved by using Algorithm 1 and Algorithm 2 under perfect CSI. The results are obtained for 1000 channel realizations. It is seen from Table 3 that Algorithm 2 can provide rank-one solutions, while Algorithm 1 cannot. The reason is that a cost function related to rank-one solutions is applied in Algorithm 2. When rank-one solutions are achieved, the optimal beamforming vectors can be obtained for the CBS and PBS; otherwise, the optimal beamforming vectors cannot be obtained. This is the reason why Algorithm 2 performs better than Algorithm 1.

\begin{figure}[!t]
\centering
\includegraphics[width=3.2 in]{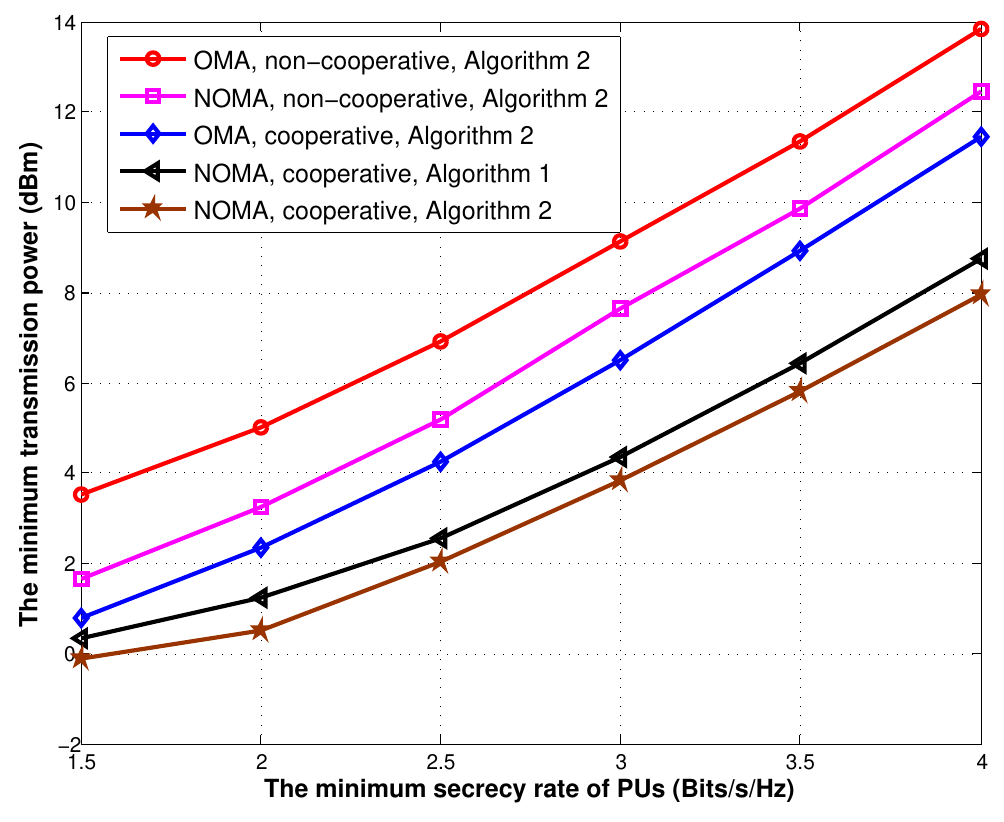}
\caption{The minimum transmission power versus the secrecy rate of PUs under perfect CSI scenario.} \label{fig.1}
\end{figure}
Fig. 3 is provided for further verifying the efficiency of the proposed cooperative jamming scheme. The number of EHRs in the secondary network is 2. It is observed that the transmission power increases with the secrecy rate of PUs. It can be readily explained by the fact that a high transmission power is required for guaranteeing the increased secrecy rate requirement of PUs. As shown in Fig. 3, the minimum transmission power required when the proposed cooperative jamming scheme is applied is lower than that in the absence of cooperation between the primary and the secondary network. This phenomenon further demonstrates the efficiency of our proposed cooperative jamming scheme for achieving secure communications. It is also seen from Fig. 3 that NOMA outperforms TDMA and that Algorithm 2 performs better than Algorithm 1 in terms of the required transmission power.

\begin{figure}[!t]
\centering
\includegraphics[width=3.2 in]{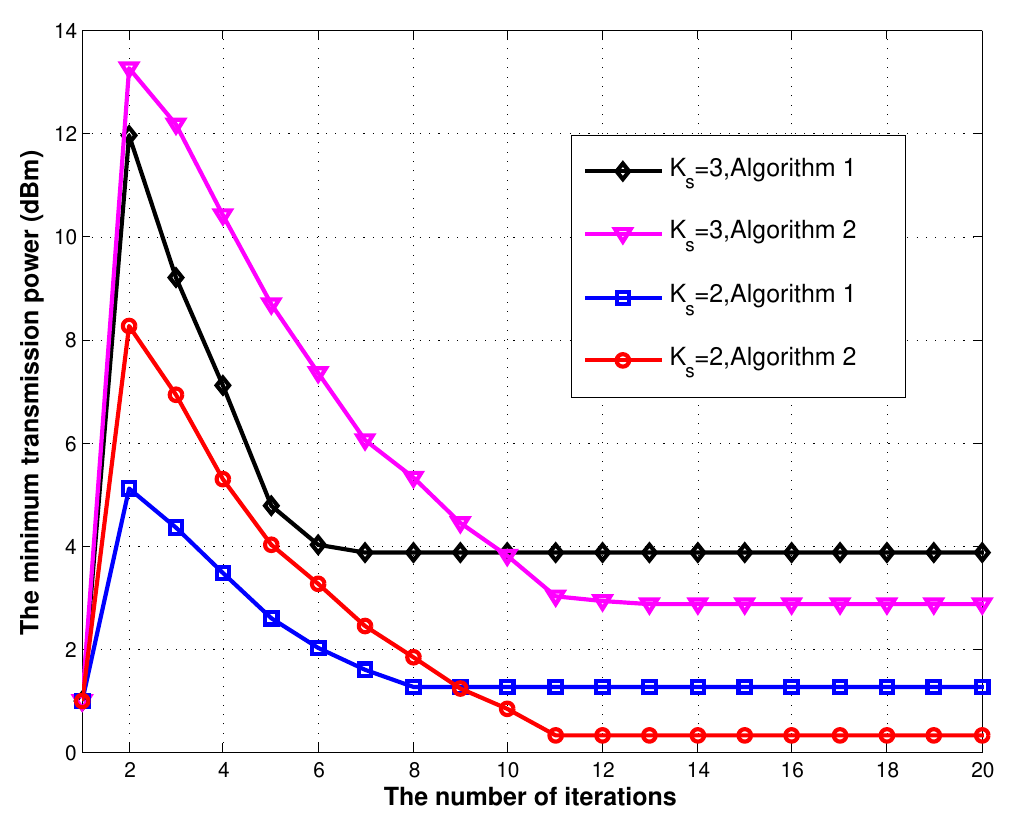}
\caption{The minimum transmission power versus the number of iterations required by using Algorithms 1  and  2.} \label{fig.1}
\end{figure}
Fig. 4 shows the minimum transmission power versus the number of iterations required by using Algorithm 1 and Algorithm 2. The number of EHRs in the secondary network is set to $2$ or $3$. It is seen from Fig. 4 that both Algorithm 1 and Algorithm 2  require only a few iterations to converge. This confirms the efficiency of our proposed algorithms. It is also seen that the number of iterations required by Algorithm 2 is larger than that required by Algorithm 1. The reason is that the complexity of Algorithm 2 is higher than that of Algorithm 1. From Fig. 3 and Fig. 4, it is seen that there is a tradeoff between the complexity of algorithms and the transmission power obtained by using these algorithms.

\begin{figure}[!t]
\centering
\includegraphics[width=3.2 in]{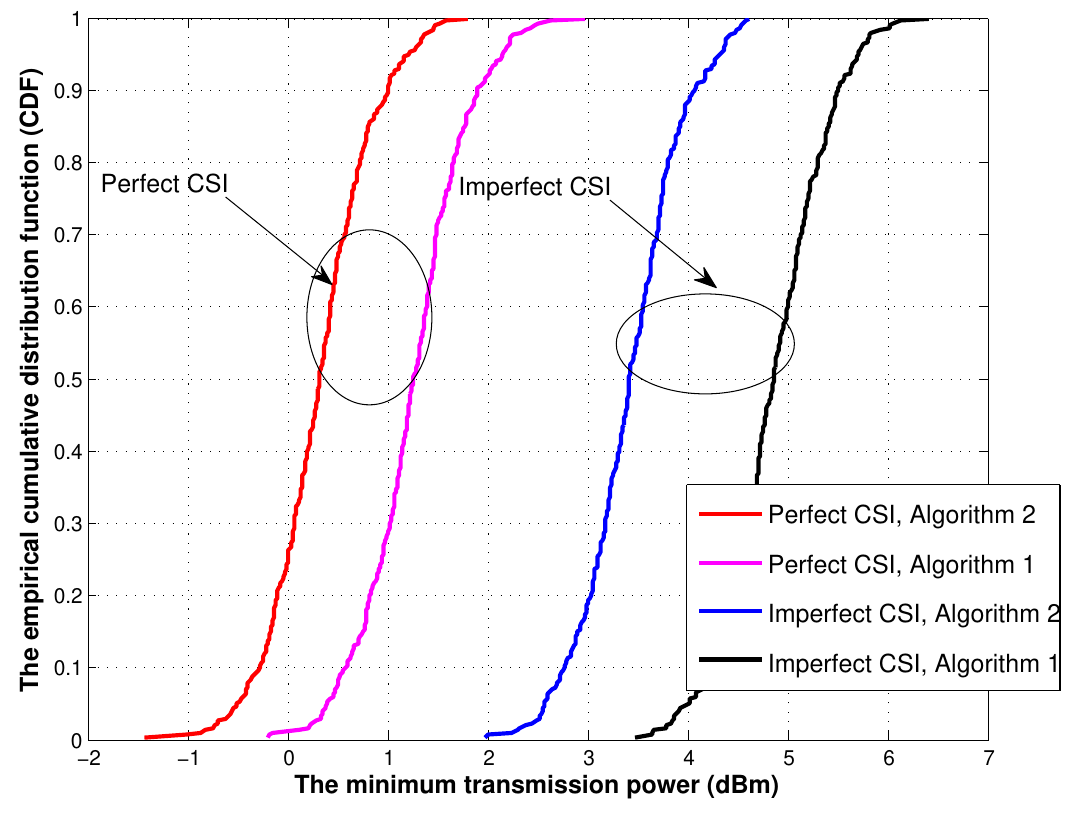}
\caption{The empirical CDF of the minimum transmission power of the CBS for both perfect and imperfect CSI scenarios.} \label{fig.1}
\end{figure}
Fig. 5 shows the empirical cumulative distribution function (CDF) of the minimum transmission power under the perfect CSI and the bounded CSI error model. The number of EHRs in the secondary network is 2. It is seen from Fig. 5 that the transmission power consumed under the bounded CSI error model is higher than that consumed under the perfect CSI, for both Algorithm 1 and Algorithm 2. The reason is that a higher transmission power is required for guaranteeing the secrecy rates of the PUs and the SUs, when the CSI is imperfect. It is also seen that Algorithm 2 is superior to Algorithm 1 in terms of the transmission power minimization, even when the CSI is imperfect.

\begin{figure}[!t]
\centering
\includegraphics[width=3.2 in]{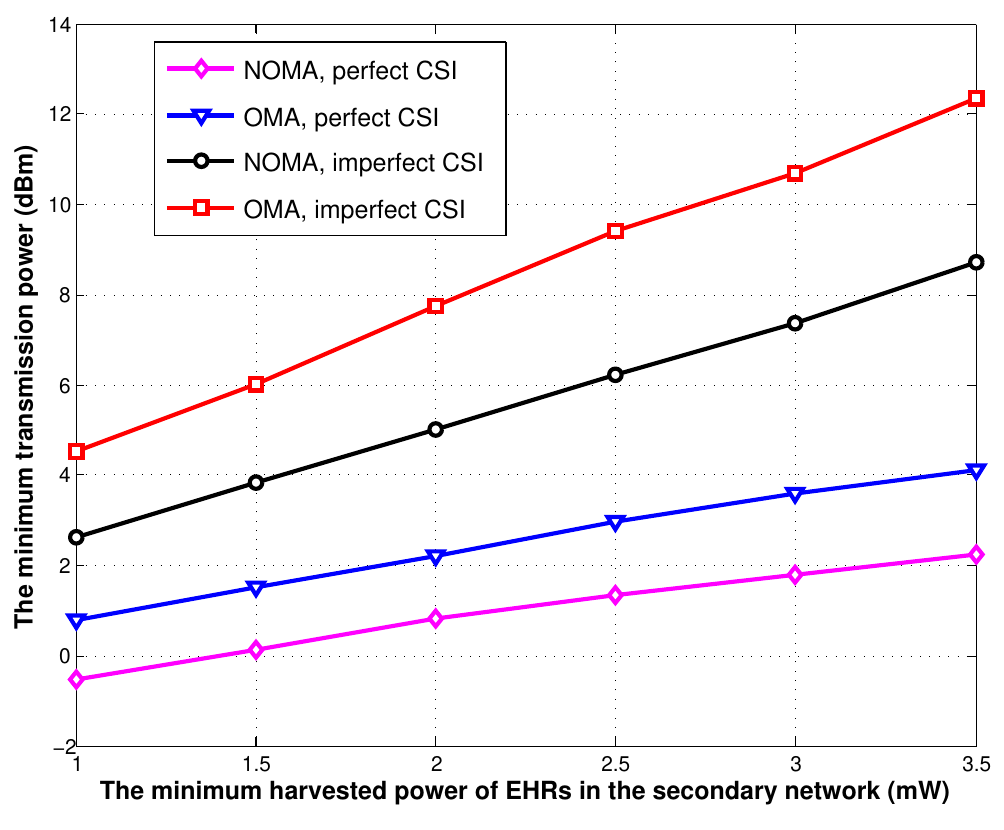}
\caption{The minimum transmission power versus the minimum harvested power requirement of EHRs in the secondary network for both perfect and imperfect CSI scenarios.} \label{fig.1}
\end{figure}
Fig. 6 shows the minimum transmission power versus the minimum harvested power requirement of EHRs in the secondary network for both perfect and imperfect CSI scenarios. The number of EHRs in the secondary network is set to 3. The simulation results are obtained by Algorithm 2. It is seen that the transmission power consumed by using TDMA is higher than that consumed by using NOMA, for both perfect and imperfect CSI scenarios. The reason is that the secrecy rate of SUs achieved by using NOMA can be higher than that obtained by using TDMA \cite{Y. Zhang1}. A lower transmission power is required to grantee the secrecy rate of SUs. It is also seen that the imperfect CSI has a significant effect on the minimum transmission power. Moreover, as shown in Fig. 6, the minimum transmission power increases with the harvested power of EHRs in the secondary network. It can be readily explained by the fact that a higher transmission power is required to satisfy the increased harvesting power of EHRs in the secondary network.

\section{Conclusions}
Secure communication was studied for MISO NOMA CRNs using SWIPT, where a practical non-linear EH model was applied.  To enhance the security of the primary network, an artificial-noise-aided cooperative jamming scheme was proposed. The transmission beamforming vectors and AN-aided covariance matrix were jointly optimized to minimize the total transmission power of the network, while the secrecy rates of both PUs and SUs as well as the EH requirement of EHRs were satisfied. The beamforming design problems were investigated under both the perfect CSI and the bounded CSI error model. A pair of algorithms were proposed to solve these challenging non-convex problems. It was shown that the performance achieved by using NOMA is better than that obtained by using OMA. Simulation results also show that the algorithm based on the cost function is superior to the algorithm based on SDR. Moreover, our proposed cooperative jamming scheme is efficient to improve the security of MISO NOMA CRNs using SWIPT.

\end{document}